\documentclass[%
 reprint,
superscriptaddress,
 amsmath,amssymb,
 aps,
]{revtex4-2}

\usepackage{graphicx, booktabs, svg}
\usepackage{dcolumn}
\usepackage{bm}
\usepackage{hyperref}

\usepackage[british]{babel}

\DeclareMathAlphabet{\mathpzc}{OT1}{pzc}{m}{it}
 
\def\beq{\begin{equation}}
\def\eeq{\end{equation}}
\def\bea{\begin{eqnarray}}
\def\eea{\end{eqnarray}}
\def\bwt{\begin{widetext}}
\def\ewt{\end{widetext}}
\def\nin{\noindent} 
\def\nn{\nonumber\\}

\begin{document}


\title{Hubble tension in k-essence: Evidence for robust tension alleviation}



\author{Isaac Opio}\email{iopio2567@gmail.com}
\affiliation{Department of Physics \& Astronomy, Botswana International University of Science and Technology, Palapye, Botswana}

\author{Didam G.A. Duniya}\email{duniyaa@biust.ac.bw}
\affiliation{Department of Physics \& Astronomy, Botswana International University of Science and Technology, Palapye, Botswana}

\author{Bishop Mongwane}\email{bishop.mongwane@uct.ac.za}
\affiliation{Department of Mathematics \& Applied Mathematics, University of Cape Town, South Africa}

\author{Hassan Abdalla}\email{hassanahh@gmail.com}
\affiliation{Centre for Space Research, North-West University, Potchefstroom 2520, South Africa}
\affiliation{Department of Astronomy and Meteorology, Omdurman Islamic University, Omdurman 382, Sudan}

\date{\today}

\begin{abstract}\nin
The Hubble tension has come to stay as a major problem in modern cosmology as it continues to plague the standard cosmological model ($\Lambda$CDM). As one of the viable, self-consistent dark energy theories, k-essence involves nontrivial self-interactions that can modify the background expansion beyond recombination; thereby impacting the sound horizon to last scattering, and hence, the inferred value of the Hubble constant. We examine this tension in two physically motivated k-essence models, dilaton and tachyon, using datasets from Planck and late-Universe probes including Pantheon+SH0ES, cosmic chronometer (CC), Supernova Cosmology Project Union compilation (Union3), Dark Energy Survey Year~5 (DESY5), and Dark Energy Spectroscopic Instrument (DESI) measurements. While $\Lambda$CDM exhibits inconsistent tension inferences, both k-essence models exhibit a substantial tension alleviation that is robust against the inclusion of the independent late-Universe cosmological datasets, giving consistent tension reduction irrespective of whether the observations are supernovae (Pantheon+SH0ES, Union3, DESY5) alone or in combination with cosmic chronometers (CC) and baryon acoustic oscillation measurements (DESI). The combined late-Universe dataset leads to only $0.14\sigma$ and $0.69\sigma$ offsets from the Planck prediction in the dilaton and tachyon models, respectively, compared to $5.89\sigma$ tension in $\Lambda$CDM. Both models demonstrate that the inferred tension alleviation is a stable, intrinsic consequence of the underlying k-essence dynamics rather than of model fine tuning: model parameters remain unchanged across datasets. The results establish that the apparent Hubble tension is not an unavoidable feature of late-Universe cosmology but depends critically on the description of dark energy. Moreover, while this analysis demonstrates resolution of specifically the Planck--late-Universe Hubble tension in k-essence cosmologies, other early-Universe probes will need to be included for resolution of the Hubble tension in general. 

\vspace{0.5cm}
\end{abstract}

\maketitle


\section{Introduction}
\label{sec:intro}
As the precision of cosmological measurements continues to sharpen, new discrepancies begin to surface in the estimation of the observables. One of such discrepancies is the inconsistency in current determinations of the present-day background expansion rate of the Universe, which is parameterized by the Hubble constant $H_0$. Particularly, within the standard cosmological model ($\Lambda$CDM), measurements based on the cosmic microwave background (CMB) radiation temperature anisotropies determine $H_0 \,{=}\, 67.4 \pm 0.5 \, {\rm km\,s^{-1}\,Mpc^{-1}}$ (Planck~2018) \cite{Planck:2018vyg}, while local determinations with extragalactic cepheid Type Ia supernovae (SNIa) obtained a significantly larger estimate with a baseline value, $H_0 \,{=}\, 73.04 \pm 1.04 \, {\rm km\,s^{-1}\,Mpc^{-1}}$ (SH0ES: Supernovae and H0 for the Equation of State of dark energy) \cite{Riess:2021jrx}. This gives an absolute difference $\Delta{H_0} \,{\sim}\, 5\sigma$, with $\sigma$ being the combined uncertainty. Separate measurements, including megamasers~\cite{Braatz:2010sg} and time delay~\cite{H0LiCOW:2016xpx}, have reinforced the existence of this discrepancy: now known as the ``Hubble tension'' \cite{DiValentino:2021izs, Hu:2023jqc, CosmoVerseNetwork:2025alb, Cai:2026swf, Hu:2023eyf, Riess:2019qba, Verde:2019ivm, Riess:2024vfa, Riess:2025chq, Pantos:2026cxv, Knox:2019rjx, Poulin:2018cxd, Abdalla:2022yfr, Copeland:2023zqz, HosseiniMansoori:2024pdq, Hussain:2024qrd, Doran:2006kp, Niedermann:2019olb, Agrawal:2019lmo, Hill:2020osr, Yang:2018euj, Bhattacharyya:2018fwb, Kumar:2019wfs, DiValentino:2019jae, Nunes:2018xbm, Escamilla-Rivera:2019ulu, Odintsov:2020qzd, Braglia:2020auw, Shimon:2020mvl, Gangopadhyay:2022bsh, Mandal:2023bzo, Petronikolou:2023cwu, Kavya:2025vsj, Verma:2025lmr, Li:2019yem, DiValentino:2020naf, Yang:2021eud, Adi:2025hyj, Zhang:2025lam, Lyu:2020lps,DiValentino:2021rjj,Yarahmadi:2025vvg}  (see \cite{DiValentino:2021izs, Hu:2023jqc, CosmoVerseNetwork:2025alb, Cai:2026swf} for reviews). Apparently, the Hubble tension is proving persistent, and now exceeding the $5\sigma$ level depending on the datasets considered (e.g.~\cite{Cai:2026swf, DiValentino:2021izs, Hu:2023jqc, CosmoVerseNetwork:2025alb, Hu:2023eyf, Riess:2019qba, Verde:2019ivm, Riess:2024vfa, Riess:2025chq, Pantos:2026cxv}). However, a brief inspection of the $H_0$ estimations in the literature indicates that the growth in tension (in $\Lambda$CDM) is almost entirely owing to uncertainty shrinking, not central values diverging. 

The standard estimation of $H_0$ from the CMB is done by matching the angular size, $\theta_* \,{=}\, r_s(z_*)/D_M(z_*)$, of the acoustic fluctuations at last scattering surface to the observed CMB power spectrum---where $r_s(z_*)$ is the comoving sound horizon to last scattering surface, at redshift $z \,{=}\, z_*$, and $D_M(z_*)$ is the comoving distance. Thus, in order to resolve the tension in $H_0$ it implies it will require (i) reducing the calculated distance sound travelled before recombination, i.e.~modifying the pre-recombination physics to reduce $r_s(z_*)$, (ii) modifying how the Universe expanded at late times to match the observed angular size of the CMB, i.e.~altering the late-time expansion history to change $D_M(z_*)$, or (iii) abandoning the standard cosmological framework in favour of new physics, possibly introducing more radical departures from $\Lambda$CDM (see e.g.~\cite{DiValentino:2021izs, Hu:2023jqc, CosmoVerseNetwork:2025alb, Knox:2019rjx, Poulin:2018cxd, Abdalla:2022yfr, Copeland:2023zqz, HosseiniMansoori:2024pdq, Hussain:2024qrd, Cai:2026swf}). Although a variety of theoretical frameworks have been explored, ranging from early dark energy (e.g.~\cite{Poulin:2018cxd, Doran:2006kp, Niedermann:2019olb, Agrawal:2019lmo, Hill:2020osr}) and interacting dark sectors (e.g.~\cite{Yang:2018euj, Bhattacharyya:2018fwb, Kumar:2019wfs, DiValentino:2019jae}) to modified gravity (e.g.~\cite{Nunes:2018xbm, Escamilla-Rivera:2019ulu, Odintsov:2020qzd, Braglia:2020auw, Shimon:2020mvl, Gangopadhyay:2022bsh, Mandal:2023bzo, Petronikolou:2023cwu, Kavya:2025vsj, Verma:2025lmr}), dynamical dark energy (e.g.~\cite{Li:2019yem, DiValentino:2020naf, Yang:2021eud, Adi:2025hyj, Zhang:2025lam}), and exotic neutrinos (e.g.~\cite{Lyu:2020lps,DiValentino:2021rjj,Yarahmadi:2025vvg}), the search for a single robust and complete solution that simultaneously satisfies the full spectrum of early-Universe probes and those of the late Universe independently, is still ongoing.

An attractive class of models with the potential to address the Hubble tension is ``k-essence" \cite{Armendariz-Picon:1999hyi, Armendariz-Picon:2000nqq, Chiba:1999ka, Armendariz-Picon:2000ulo, Piazza:2004df, Amendola:2010bk, Duniya:2026aeu, Bagla:2002yn, Chimento:2003ta, Copeland:2023zqz, HosseiniMansoori:2024pdq, Hussain:2024qrd}, which involves scalar-field $\phi$ dynamics beyond canonical quintessence \cite{Peebles:1987ek, Brax:1999gp, Barreiro:1999zs, Amendola:2010bk, Tsujikawa:2013fta}. The k-essence theory, a scalar-field theory with non-canonical kinetic structure, provides a flexible and theoretically motivated extension of standard dark energy. (Originally studied for inflation \cite{Armendariz-Picon:1999hyi}; later, for late-time acceleration \cite{Armendariz-Picon:2000nqq, Chiba:1999ka, Armendariz-Picon:2000ulo, Chimento:2003ta, Duniya:2026aeu}). In this theory the Lagrangian depends non-linearly on the kinetic term $X \,{\propto}\, \nabla_\mu\phi\nabla^\mu\phi$, leading to rich phenomenology including a variable sound speed (different from unity), non-trivial attractor solutions, and a dynamical equation of state parameter. Unlike quintessence, k-essence allows for non-trivial kinetic self-interactions that can modify the background expansion beyond recombination by contributing to the total energy density, impacting $r_s(z_*)$ and hence the value of $H_0$.

While several extensions to $\Lambda$CDM, including k-essence, have been reported in the literature to alleviate the Hubble tension, a lot of these studies are based on a joint analysis of the early- and late-Universe datasets (see e.g.~\cite{Copeland:2023zqz, HosseiniMansoori:2024pdq, Hussain:2024qrd}, for the case of k-essence). Such approaches provide complementary insights rather than a direct test of the tension, which requires:
\begin{enumerate}
\item Estimating two independent constraints within the same model from
\begin{itemize}
\item early-time data (e.g.~Planck), $H_0^{\rm early}$, and
\item late-time data (e.g.~SH0ES), $H_0^{\rm late}$.
\end{itemize}
\item Explicitly computing $\Delta{H_0} = |H_0^{\rm late}- H_0^{\rm early}|$, the discrepancy between the two in-model inferences.
\end{enumerate}
Essentially, a direct test involves independent parameter inference from early- and late-Universe probes within the same model, followed by assessment of direct discrepancy. This matters because the Hubble tension is fundamentally a disagreement between independently inferred values from disjoint redshift $z$ regimes ($z \,{\gtrsim}\, 10^3$ and $z \,{\lesssim}\, 3$). 

Joint fits can mask the tension e.g. by producing an apparently favoured intermediate $H_0$ value while still leaving the individual estimates from each dataset separated beyond consistency. It has become standard in physics to use a confidence level ${>}\, 99\%$ (which permits $1\%$ chance of statistical error) to assess measurement discrepancies. Typically, a difference of $3\sigma$ ($99.73\%$ confidence) is the minimum level required to call a discrepancy a tension, with $5\sigma$ ($99.99994\%$ confidence) being the ``gold standard" for definitive tension. Thus, for a tension to be considered ``resolved," the discrepancy needs to be well below the $3\sigma$ threshold. If $\Delta{H_0} \,{<}\, 3\sigma$, statistically the discrepancy is no longer considered a tension; at $\Delta{H_0} \,{\leq}\, 1\sigma$, the error bars overlap completely, and the two estimates are statistically consistent or in direct agreement. Majority of the joint analyses in the literature show alleviation in a joint-fit sense, but do not rigorously demonstrate direct agreement. A joint fit by itself is not sufficient to resolve the Hubble tension in a given model, unless independent estimates from the disjoint $z$ regimes are determined to be in direct agreement within the model.

In this paper we investigate, by direct test, the ability of k-essence to resolve the Hubble tension between Planck and late-Universe probes, while satisfying current cosmological constraints. We consider two physically motivated k-essence models: dilatonic ghost condensate and tachyon field (e.g.~\cite{Piazza:2004df, Amendola:2010bk, Bagla:2002yn, Chimento:2003ta, Duniya:2026aeu}). We probe the allowed parameter space in which k-essence dynamics can meaningfully alter the inferred value of $H_0$. The rest of this paper is organized \emph{viz.}: In \S\ref{sec:k-essence} we discuss the background equations for a universe dominated by radiation, baryons, cold dark matter and k-essence. We discuss the datasets and methods considered, in \S\ref{sec:data}. We present and discuss our results in \S\ref{sec:results}, and conclude in \S\ref{sec:conclusion}.

\section{Background Universe with K-essence}
\label{sec:k-essence}
In this work we take the Universe as being dominated by radiation (relativistic particles), baryons, cold dark matter, and k-essence $\phi$.

\subsection{The Background Equations}
\label{subsec:ConsvEqs}
In a universe dominated by radiation ($A \,{=}\, r$), baryons ($A \,{=}\, b$), cold dark matter ($A \,{=}\, c$), and k-essence ($A \,{=}\, \phi$), the conservation of total energy-momentum tensor gives
\beq\label{gen-conservation-Eqn}
\dot{\rho}_A + 3H(1+w_A)\rho_A = 0,\quad w_A = P_A/\rho_A
\eeq
where a dot denotes time derivative with respect to physical time, $H \,{=}\, \dot{a}/a$ is the physical Hubble parameter, $a \,{=}\, 1/(1+z)$ is the spacetime scale factor, $\rho_A$ and $P_A$ are the energy desnsity and pressure for $A$, respectively, and $w_A$ are the equation-of-state parameters:
\beq\label{EoSs}
w_r = 1/3,\quad w_b = w_c = 0,
\eeq
with $P_\phi$ is to be specified ($w_\phi \,{=}\, P_\phi/\rho_\rho$), and 
\beq\label{rhophi-X}
\rho_\phi = 2X\partial_X P_\phi - P_\phi,\quad X \equiv -\dfrac{1}{2}\nabla^\mu\phi \nabla_\mu\phi = \dfrac{\dot{\phi}^2}{2}.
\eeq
The solution to the general background conservation equation \eqref{gen-conservation-Eqn}, is given by
\bea\label{rho_A}
\rho_A &=& \rho_{A0}\, \exp\left[3\int^1_a \left(1+w_A(\tilde{a})\right)\frac{d\tilde{a}}{\tilde{a}}\right], \nn
&\equiv & \rho_{A0}\, \exp\left[f_A(a)\right],
\eea
where $\rho_{A0}$ are the energy densities at today ($z \,{=}\, 0$). 

Thus, the Friedmann equation (in flat homogeneous and isotropic metric) becomes
\beq\label{Friedmann}
\dfrac{{H}^2(a)}{H^2_0} = \dfrac{\Omega_{r0}}{a^{4}} + \dfrac{\Omega_{m0}}{ a^{3}} 
+ \Omega_{\phi 0}\, e^{f_\phi(a)} \equiv E^2(a),
\eeq
where $\Omega_{A0} \,{=}\, \kappa^2\rho_{A0}/(3H_0^2)$ are the present-day values of energy density parameters for the cosmic species $A$, $\kappa \,{\equiv}\, \sqrt{8\pi{G}}$ with $G$ being the Newton's gravitational constant, $H_0$ is the Hubble constant (present-day value of the Hubble parameter), $f_\phi$ is as in \eqref{rho_A}, and
\beq\label{Omega_rm_eq}
\Omega_{r0} = \dfrac{\Omega_{m0}}{1+z_{\rm eq}},\quad z_{\rm eq} =  \dfrac{2.5 \times 10^4}{\left(T_{\rm CMB}/2.7K\right)^4} \Omega_{m0}h^2,
\eeq
with $\Omega_{m0}$ being the matter (baryons plus cold dark matter) density parameter, $z_{\rm eq}$ being the redshift at radiation-matter equality, $T_{\rm CMB} \,{\approx}\, 2.7255K$ \cite{Chen:2018dbv} is the mean value of the CMB radiation temperature, and 
\beq\label{h}
h \equiv \dfrac{H_0}{100\, \rm km/s/Mpc},
\eeq
is the reduced Hubble constant (dimensionless), with $H_0$ being in km/s/Mpc.

Evaluating \eqref{Friedmann} will require $w_\phi$, which in turn will require the evolutions of $\phi$ and $\dot{\phi}$ that must come from solving the k-essence equation of motion: 
\beq\label{Klein-Gordon}
\ddot{\phi} + 3 H \Gamma (\partial_X P_\phi) \dot{\phi} + \Gamma (2X \partial^2_{X\phi} P_\phi - \partial_\phi P_\phi) = 0,
\eeq
where we used \eqref{gen-conservation-Eqn} and \eqref{rhophi-X}, $X$ is as in \eqref{rhophi-X}, and 
\beq\label{Gamma}
\Gamma^{-1} \equiv \partial_X P_\phi + 2X \partial^2_X P_\phi,
\eeq
with $\partial^2_{X\phi}$ denoting successive partial derivatives with respect to $X$ and $\phi$, respectively (similarly for $\partial_\phi$). To analyse \eqref{Friedmann}, we need to specify an exact form for $P_\phi$. Thus, we consider the dilatonic ghost condensate (``dilaton,'' henceforth) and tachyon models (e.g.~\cite{Piazza:2004df, Amendola:2010bk, Bagla:2002yn, Chimento:2003ta, Duniya:2026aeu}).


\subsection{Dilaton model}
\label{subsec:Dilaton}

The dilatonic model \cite{Piazza:2004df, Amendola:2010bk, Duniya:2026aeu} is described by the Lagrangian (density), given by
\begin{equation}\label{Dilaton}
P_\phi = -X + \dfrac{X^2}{M^8} U(\phi),
\end{equation}
where $X$ is as in \eqref{rhophi-X}, $M$ is a (constant) mass scale, and
\beq\label{U-dil}
U(\phi) = M^4e^{\lambda\kappa\phi},
\eeq
is the potential, with $\lambda$ being a free (dimensionless) constant and, $\kappa$ is as in \eqref{Friedmann}. By using \eqref{rhophi-X} and \eqref{Dilaton} we obtain the dilaton particular equation-of-state parameter:
\beq\label{w-dil}
w_\phi = \dfrac{1 - X U(\phi)/M^8}{1 - 3X U(\phi)/M^8} , 
\eeq
where the behaviour of $w_\phi$ is governed by both the Klein-Gordon equation \eqref{Klein-Gordon} and the potential \eqref{U-dil}. 


\subsection{Tachyon model}
\label{subsec:Tachyon}
The tachyon field \cite{Amendola:2010bk, Bagla:2002yn, Chimento:2003ta, Duniya:2026aeu} has a Lagrangian, given by 
\beq\label{Tachyon}
P_\phi = -U(\phi)\sqrt{1 - \dfrac{2X}{M^4}},
\eeq
where $M$ is a in \eqref{Dilaton}, and the potential is 
\beq\label{U-tach}
U(\phi) = \dfrac{M^{4+\alpha}}{\phi^\alpha},\quad \alpha > 0,
\eeq
with $\alpha$ being a constant (dimensionless). Using \eqref{rhophi-X} and \eqref{Tachyon}, we get the tachyon particular $w_\phi$, given by
\beq\label{w-tach}
w_\phi = -1 + \dfrac{2X}{M^4}, 
\eeq
where we note that for tachyon, the equation of state parameter is not directly connected to the potential \eqref{U-tach}, unlike in the case of dilaton \eqref{w-dil}. (See e.g.~\cite{Amendola:2010bk, Duniya:2026aeu}, for the full particular evolution equations of both models.) 

Unlike in canonical quintessence, we see by \eqref{Tachyon} and the second term in \eqref{Dilaton} that, k-essence carries nontrivial self-interactions. These can impact the value of $H_0$ by modifying the expansion history directly via $\rho_\phi$ or $\Omega_\phi$, and indirectly via $w_\phi$.


\section{Observational Analysis}\label{sec:data}
\subsection{SNIa Analysis}\label{subsec:SNIa}

We use the Pantheon+SH0ES dataset~\cite{Scolnic:2021amr, Brout:2022vxf, PantheonPlusSH0ESData}, which is basically the SH0ES-calibrated Pantheon+ dataset, with sample in the redshift range $0.001 \,{<}\, z\,{<}\, 2.26$. Pantheon+SH0ES combines both precision supernova distances and absolute calibration, providing sharper late-Universe measurement of $H_0$, better than either SH0ES or Pantheon+ alone. We also use Union3 sample (spanning $0.05 \,{<}\, z\,{<}\, 2.26$) \cite{Rubin:2023jdq}, and the five-year dataset of the Dark Energy Survey (DESY5) \cite{DES:2024jxu} with combined sample in the redshift range $0.025 \,{<}\, z \,{<}\, 1.13$. The Pantheon+SH0ES and Union3 datasets provide the standardised or corrected SNIa apparent magnitudes $m_{\rm B}$ (see e.g.~\cite{Scolnic:2021amr, Riess:2021jrx, Brout:2022vxf}) and, both cosmological distance moduli (for Hubble-flow SNIa) and independently measured cepheid distance moduli (for calibrator-host SNIa); whereas, the DESY5 dataset provides mainly cosmological SNIa distance moduli.

For Pantheon+SH0ES and Union3 we fit $m_{\rm B}$ rather than the pre-computed distance moduli. Hence, the (standardised) absolute magnitude $M_{\rm B}$ is taken as a free, calibration parameter. For Hubble-flow SNIa, which lie in the cosmological expansion regime, the predicted or theoretical apparent magnitude, is given by
\beq\label{m_th}
m^{\rm th}_{\rm B}(z) = \mu_{\rm Cosmo}(z) + M_{\rm B},
\eeq
where $\mu_{\rm Cosmo}$ is the distance modulus predicted by the underlying background cosmology, given by
\beq\label{mu_th}
\mu_{\rm Cosmo}(z) = 25 + 5\log_{10}\left[\dfrac{D_L(z)}{\rm Mpc}\right],
\eeq
with $D_L$ being the cosmological luminosity distance:
\beq\label{D_L}
D_L(z) = \dfrac{c}{H_0}(1+z) \int_{0}^{z}\dfrac{dz'}{E(z')},
\eeq
$c$ is the speed of light (in km/s), and $E$ is as in \eqref{Friedmann}. Essentially, for these supernovae, the luminosity distance is inferred from the distance-redshift relation \eqref{D_L}.

For calibrator-host SNIa (in the local Universe), the luminosity distance is not inferred from a redshift-dependent relation, e.g.~\eqref{D_L}, but instead is determined directly from the local distance ladder, without the assumption of a cosmological expansion model: the cosmological $D_L$ to a host galaxy is replaced in the likelihood by the corresponding cepheid distance $D_{\rm Ceph}$, which directly calibrates $M_{\rm B}$. Thus, the theoretical apparent magnitude for calibrator-host SNIa is predicted by the cepheid distance modulus $\mu_{\rm Ceph}$, given by
\beq\label{m_Ceph}
m^{\rm th}_{\rm B} = \mu_{\rm Ceph} + M_{\rm B}.
\eeq
Note that although calibrator hosts do have redshifts, the distance is not obtained by inputting the redshift into a distance-redshift relation. Empirically, there is a correlation between distance and redshift for these hosts, which is explored to determine the distance directly. Thus $D_{\rm Ceph}$, and hence $\mu_{\rm Ceph}$, are given as fixed observational measurements (with associated covariance) that are inferred independently of the host redshift and therefore constitute external distance information. Consequently, the calibrator contribution to the likelihood is independent of the cosmological parameters entering $D_L(z)$.

Using Hubble-flow SNIa alone, $M_{\rm B}$ and $H_0$ are degenerate: at low redshifts ($z \,{\lesssim}\, 0.15$), where $D_L \,{\propto}\, 1/H_0$ at fixed cosmological parameters, $m_{\rm B} = 5\log_{10}(D_L/{\rm Mpc}) + M_{\rm B} + 25$ implies that low-redshift SNIa data only constrain ${\cal M} \,{\equiv}\, M_{\rm B} \,{-}\, 5\log_{10}{H_0}$, which can absorb changes in both $M_{\rm B}$ and $H_0$. However, the cepheid calibrators break this degeneracy by providing external, cosmology-independent distance information through their measured distance moduli, anchoring $M_{\rm B}$. This fixes the absolute distance scale and enables the inference of $H_0$. This is why the calibrator substitution \eqref{m_Ceph} is necessary. 

The likelihood is evaluated in $m_{\rm B}$ space in a Markov Chain Monte Carlo (MCMC) analysis, using the full covariance matrices provided by the Pantheon+SH0ES and Union3 analyses. The chi-square statistic, with $m_{\rm B}$ and $m^{\rm th}_{\rm B}$ in vector form, is given by
\beq\label{ch2_PP_U3}
\chi^2_d = \left(\boldsymbol{m}_{\rm B} - \boldsymbol{m}^{\rm th}_{\rm B}(\boldsymbol{\hat \theta})\right)^T C^{-1}_d \left(\boldsymbol{m}_{\rm B} - \boldsymbol{m}^{\rm th}_{\rm B}(\boldsymbol{\hat \theta})\right),
\eeq
where henceforth bold faces denote vectors, $C^{-1}_d$ are the inverse covariance matrices for the datasets ($d$ = Pantheon+SH0ES, Union3), and $\boldsymbol{m}^{\rm th}_{\rm B}(\boldsymbol{\hat \theta})$ in unified piecewise form for a given dataset, is given by (see also \cite{Perivolaropoulos:2023iqj})
\bea\label{m_th_comb}
m^{\rm th}_{{\rm B},i}(\boldsymbol{\hat \theta}) = 
\begin{cases} 
\mu_{{\rm Cosmo},i}(\boldsymbol{\theta}) + M_{\rm B}, & \mbox{Hubble-flow SNIa}, \\
\\
\mu_{{\rm Ceph},i} + M_{\rm B}, & \mbox{Calibrator SNIa}. 
\end{cases}
\eea
which follows from \eqref{m_th} and \eqref{m_Ceph}, with $m^{\rm th}_{{\rm B},i}$ being the prediction for the $i$-th SNIa at $z \,{=}\, z_i$ (similarly for $\mu_{{\rm Cosmo},i}$ and $\mu_{{\rm Ceph},i}$) as $i$ runs through the data values, and $\boldsymbol{\hat\theta} \,{=}\, (\boldsymbol{\theta},\, M_{\rm B})$ with $\boldsymbol{\theta}$ being the vector of common cosmological parameters (to be inferred), given by
\beq\label{theta}
\boldsymbol{\theta} = \left\lbrace H_0, \Omega_{b0}h^2, \Omega_{c0}h^2, A^D_0, A^T_0, \lambda, \alpha \right\rbrace ,
\eeq
where $M_{\rm B}$ is the same for both Hubble-flow and calibrator SNIa, $\Omega_{b0}$ and $\Omega_{c0}$ are the present-day baryon and cold dark matter density parameters, respectively, and $A^I_0 \equiv \kappa^2 M^4/(3H_0^2)$ (dimensionless) for dilaton ($I \,{=}\, D$) and tachyon ($I \,{=}\, T$); $\lambda$ and $\alpha$ are as in \eqref{U-dil} and \eqref{U-tach}. 

For DESY5 we fit the observed distance modulus $\mu$, with the chi-square statistic given by
\beq\label{ch2_DESY5}
\chi^2_{\rm DESY5} = \left(\boldsymbol{\mu} - \boldsymbol{\mu}_{\rm Cosmo}(\boldsymbol{\theta})\right)^T C^{-1}_{\rm DESY5} \left(\boldsymbol{\mu} - \boldsymbol{\mu}_{\rm Cosmo}(\boldsymbol{\theta})\right),
\eeq
where ${\mu}_{\rm Cosmo}$ and $\boldsymbol{\theta}$ are as given by \eqref{mu_th} and \eqref{theta}, respectively, and $C^{-1}_{\rm DESY5}$ is the inverse of the full DESY5 covariance matrix. We perform separate combination joint fits of Pantheon+SH0ES with Union3 and DESY5.


\subsection{CMB Analysis}
\label{subsec:CMB}
We use reduced Planck parameters from CMB analysis \cite{Chen:2018dbv, Zhai:2018vmm, Yang:2025vnm, Yang:2025oax, Wang:2013mha, Cai:2014ela, Chen:2018dbv}. In CMB measurements, the distance to the last scattering surface can be accurately determined from the locations of peaks and troughs of acoustic oscillations. 

Two important quantities with regards to the estimation of $H_0$ from the CMB are the acoustic scale $\ell_A$ and the shift parameter $R$, given by
\bea\label{l_A}
\ell_A &=& \pi \dfrac{D_M(z_*)}{r_s(z_*)},\\
\label{R}
R &=& \dfrac{H_0}{c} \sqrt{\Omega_{m0}}\, D_M(z_*),
\eea
where $z_*$ is the redshift at last scattering surface epoch, $D_M(z_*)$ is the (transverse) comoving distance at the last scattering surface, given by
\beq\label{D_M}
D_M(z) = \dfrac{D_L(z)}{1+z},
\eeq
with $D_L$ being the luminosity distance \eqref{D_L} and, $r_s(z_*)$ is the comoving sound horizon to the last scattering surface:
\beq\label{r_s}
r_s(z) = \dfrac{c}{\sqrt{3}\,H_0}\int^{1/(1+z)}_0\dfrac{da}{a^2E(a)\sqrt{1+\frac{3\Omega_{b0}h^2}{4\Omega_{r 0}h^2} a}},
\eeq
where $E(a)$ and $\Omega_{b0}$, and $\Omega_{r 0}$ are as in \eqref{Friedmann} and \eqref{Omega_rm_eq}.  

A widely used fitting formula for the redshift at last scattering surface, is given by \cite{Hu:1995en}
\beq\label{z-star}
z_* = 1048\left[1 + 0.00124(\Omega_{b0}h^2)^{-0.738}\right]\left[1 +g_1(\Omega_{b0}h^2)^{g_2}\right],
\eeq
with
\beq\nonumber
g_1 = \dfrac{0.0783\, (\Omega_{b0}h^2)^{-0.238}}{1+39.5\, (\Omega_{b0}h^2)^{0.763}},\quad g_2 = \dfrac{0.560}{1 + 21.1\, (\Omega_{b0}h^2)^{1.81}}.
\eeq
Similarly, the CMB $\chi^2$ statistic, is given by
\beq\label{chi2-Planck}
\chi^2_{\rm Planck} = \left[\boldsymbol{x} - \boldsymbol{x}_{\rm th}(\boldsymbol\theta)\right]^{\rm T}C^{-1}_{\rm Planck} \left[\boldsymbol{x} - \boldsymbol{x}_{\rm th}(\boldsymbol\theta)\right],
\eeq
where $\boldsymbol{x}$ is the vector of quantities estimated from Planck CMB measurements, given by
\beq\label{x-CMB}
\boldsymbol{x} = \left\lbrace \ell_A({\rm Planck}),\, R({\rm Planck}),\, \Omega_{b0}h^2({\rm Planck})\right\rbrace,
\eeq
and $\boldsymbol{x}_{\rm th}(\boldsymbol\theta)$ is the vector of predicted quantities \eqref{l_A}--\eqref{D_M} for the given inference parameters \eqref{theta}; with $h$ being as in \eqref{h}. The chain used for determining $\ell_A$, $R$, and $\Omega_{b0}h^2$ are obtained from one of the public Planck 2018 chains, \verb+base plikHM TTTEEE low+$\ell$ \verb+low+$E$. We use the mean values of $\ell_A(\rm Planck)$, $R(\rm Planck)$, and $\Omega_{b0}h^2({\rm Planck})$ presented by \cite{Chen:2018dbv} and their correlation matrices. 


\subsection{BAO Analysis}
\label{subsec:BAO}
We use the three-year baryon acoustic oscillation (BAO) measurements from the Dark Energy Spectroscopic Instrument (DESI) Data Release~2 (DR2) \cite{DESI:2024mwx, DESI:2025zgx, Yang:2025oax}, which contains distance ratios including $D_{\rm M}/r_{\rm d}$, $D_{\rm H}/r_{\rm d}$, and $D_{\rm V}/r_{\rm d}$, where $D_{\rm M}$ is the (transverse) comoving angular diameter distance, $D_{\rm H}$ is the radial Hubble distance, $D_{\rm V}$ is the angle-averaged distance, and $r_{\rm d}$ is the drag-epoch comoving sound horizon value.

In a homogeneous and isotropic universe, we have 
\beq\label{DH-DV}
D_{\rm H}(z) = c/H(z), \quad D_{\rm V}(z) = \left[z D_{\rm M}(z)^2 D_{\rm H}(z)\right]^{\frac{1}{3}},
\eeq
where $H$ and $D_M$ are as in \eqref{Friedmann} and \eqref{D_M}, respectively, and $r_{\rm d}$ is computed from the sound horizon \eqref{r_s} at radiation drag $z \,{=}\, z_{\rm d}$, which has a fitting formula given by \cite{Hu:1995en, Eisenstein:1997ik}	
\beq\label{z_d}
z_{\rm d} = \dfrac{1291 (\Omega_{m0}h^2)^{0.251}}{1 + 0.659 (\Omega_{m0}h^2)^{0.828}} \left[1 + b_1 (\Omega_{b0}h^2)^{b_2}\right],
\eeq
where,
\begin{align}\nonumber
b_1 =&\; 0.313\, (\Omega_{m0}h^2)^{-0.419} \left[1+0.607\,(\Omega_{m0}h^2)^{0.674}\right],\nn
b_2 =&\; 0.238\, (\Omega_{m0}h^2)^{0.223}.
\end{align}

We obtain the $\chi^2$ statistic for BAO data using
\beq
\chi^2_{\rm DESI} = \left[\boldsymbol{d}-\boldsymbol{d}_{\rm th}(\boldsymbol\theta)\right]^{\rm T} C^{-1}_{\rm DESI}\left[\boldsymbol{d}-\boldsymbol{d}_{\rm th}(\boldsymbol\theta)\right],
\eeq
where $C^{-1}_{\rm DESI}$ is the inverse covariance matrix in DR2 \cite{DESI:2025zgx}, and $\boldsymbol{d}$ is the vector of DESI observables:
\bea
\boldsymbol{d} &=& \left\lbrace D_{\rm M}/r_{\rm d}({\rm DESI}),\, D_{\rm H}/r_{\rm d}({\rm DESI}), \right.\nn
&& \left. D_{\rm V}/r_{\rm d}({\rm DESI}) \right\rbrace ,
\eea
where $\boldsymbol{d}_{\rm th}(\boldsymbol{\theta})$ is the theoretical predictions using \eqref{D_M} and \eqref{DH-DV}, accordingly; with $\boldsymbol{\theta}$ being as in \eqref{theta}. 

\begin{figure*}\centering 
\includegraphics[scale=0.55]{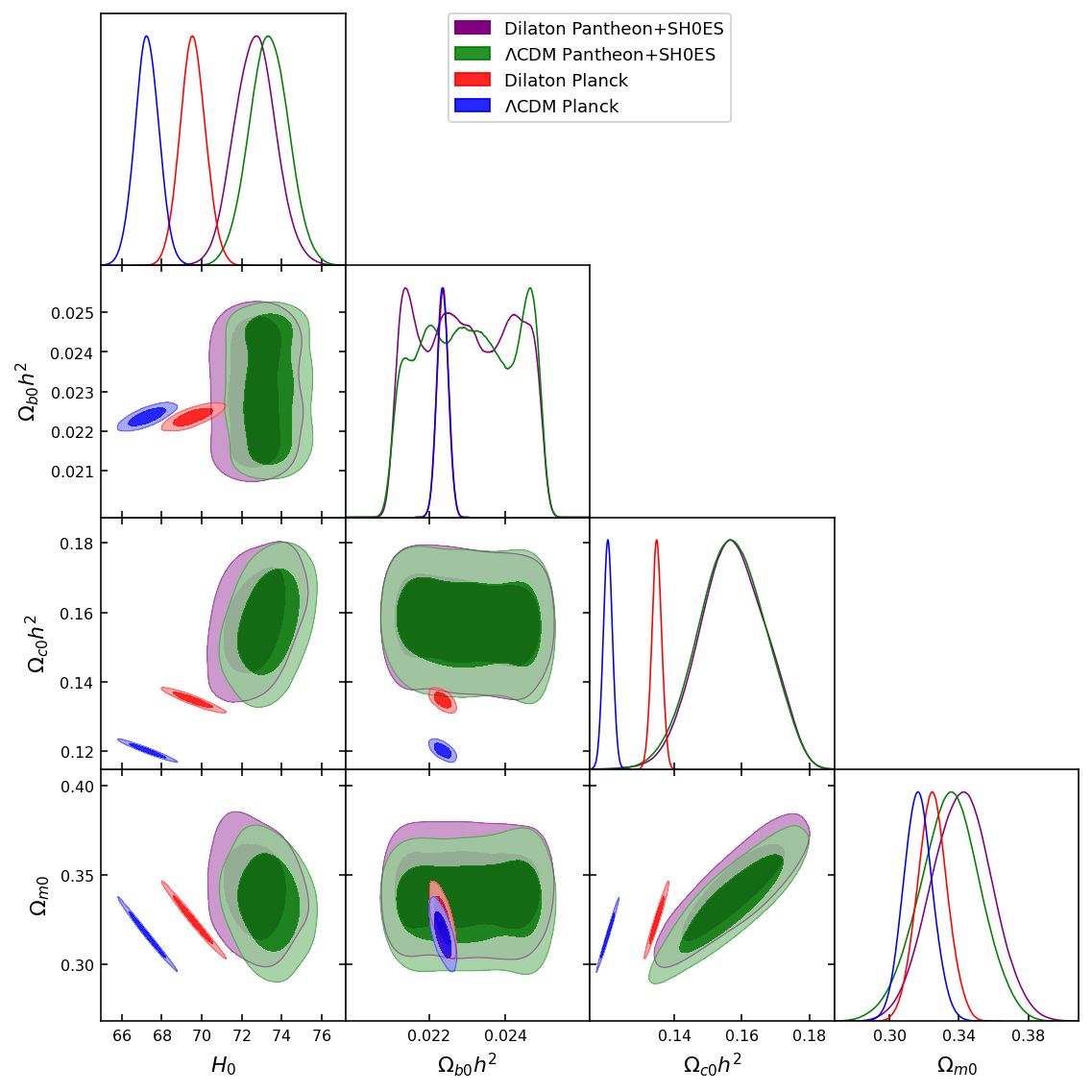}
\caption{\emph{Dilaton}: The contour plots of cosmological parameter estimates with Pantheon+SH0ES and Planck~2018 datasets, for the dilaton and cosmological-constant models. The associated probability density functions are given in the outer panels.}
\label{fig:dilaton-P18-PPS}
\end{figure*}

We also use Hubble parameter data \cite{Yu:2017iju} from cosmic-chronometric (CC) method measurements of BAO signal (in galaxy and Ly$\alpha$ forest distributions \cite{Moresco:2012jh, BOSS:2014hwf, BOSS:2016wmc}). The dataset contains measured $H(z)$ values at $0.07 \,{\leq}\, z \,{\leq}\, 2.36$. The chi-square statistic is computed by
\beq\label{chi2_CC}
\chi^2_{\rm CC} = \sum_{i}\left[\dfrac{H(z_i) - H_{\rm th}(z_i)}{\sigma_i}\right]^2,
\eeq
where $H_\text{th}(z_i)$ is theory predicted Hubble parameter \eqref{Friedmann}, and $H(z_i)$ represents the CC measured Hubble parameter value at an observed redshift $z_i$, with $\sigma_i$ uncertainty.


\section{Results and discussion}\label{sec:results}

We employ the MCMC method to determine the best-fit values and posterior distributions of all parameter sets, using initial guesses, as given in Table.~\ref{priors}. We assume $\Lambda$CDM-like initial conditions for the background cosmological evolutions in the k-essence models. 

\begin{table}[!ht]\centering
	\caption{The prior range of model parameters.}
	\label{priors}
	\begin{tabular}{lr}
		\toprule
		Parameter							&  Range\\
		\midrule
		$H_0$ 						& $[60.0,\ 80.0]$\\
		$\Omega_{b0}h^2$ 			& $[0.02,\ 0.03]$\\
		$\Omega_{c0}h^2$ 			& $[0.1,\ 0.2]$\\
		$\lambda$ 					& $ [0.01,\ 0.02]$\\
		$\alpha$ 					& $ [0.02,\ 0.03]$\\
		$A^D_{0}$ 	& $ [0.5,\ 1.5]$\\
		$A^T_{0}$ 	& $ [0.8,\ 2.0]$\\
		\bottomrule
	\end{tabular}
\end{table}

\subsection{Planck--Pantheon+SH0ES $H_0$ Tension}
\label{subsec:Planck_PP}
We first reevaluate the standard Hubble tension, albeit here between Planck and Pantheon+SH0ES, rather than specifically with SH0ES alone. We note that Pantheon+SH0ES relatively has better statistical precision, improved control of systematics, and a wider data coverage that encompasses the range of SH0ES. Thus, here we focus primarily on $H_0$ estimates from Planck~2018 and Pantheon+SH0ES; performing our analyses in both $\Lambda$CDM and the k-essence models (\S\ref{subsec:ConsvEqs} -- \S\ref{subsec:Tachyon}). 

In Fig.~\ref{fig:dilaton-P18-PPS} we show the contour plots of the cosmological constraints for the dilaton model and $\Lambda$CDM, using Planck and Pantheon+SH0ES. We see that the dominant change is the noticeable shift in the Planck $H_0$ prediction in the dilaton model, accompanied by a relatively larger $\Omega_{c0}h^2$ value. For the Pantheon+SH0ES constraints, the shifts are statistically small, with the data preferring the cosmology staying largely the same. We summarise the numerical shifts in the mean or best-fit values of the cosmological parameters (dilaton relative to $\Lambda$CDM), in Table~\ref{tab:dilaton_mean_comparisons}. In the dilaton model, the Planck best-fit value moves upward in $H_0$ ($+2.30\, {\rm km s^{-1} Mpc^{-1}}$) and that of Pantheon+SH0ES moves slightly downward (${<}\, 1\, {\rm km s^{-1} Mpc^{-1}}$); whereas, for the density parameters, while $\Omega_{b0}h^2$ remains unchanged (negligible shift, relative to the error) for both Planck and Pantheon+SH0ES, there is an upward shift in $\Omega_{c0}h^2$ ($+0.0145$) for Planck with little to no change for Pantheon+SH0ES.

For the actual Hubble tension inference, the Planck and Pantheon+SH0ES datasets exhibit a significant disagreement within $\Lambda$CDM, as already established in the literature. We have that while Planck predicts a present-day expansion rate $H_0 \,{=}\, 67.27 \,{\pm}\, 0.6\, {\rm km\,s^{-1} Mpc^{-1}}$, Pantheon+SH0ES prefers $H_0 \,{=}\, 73.37 \, {\pm}\, 0.98$ ${\rm km\, s^{-1} Mpc^{-1}}$, resulting in a tension of $5.31\sigma$. This reflects the persistent mismatch between early- and late-Universe probes and further highlights existing evidence in the literature that $\Lambda$CDM struggles to simultaneously accommodate both measurements. However, k-essence appears to substantially reduce this tension. The dilaton model alleviates the tension by producing $H_0 \,{=}\, 69.57 \,{\pm}\, 0.64\, {\rm km\,s^{-1} Mpc^{-1}}$ from Planck and $H_0 \,{=}\, 72.70 \,{\pm}\, 1.00\, {\rm km\,s^{-1} Mpc^{-1}}$ from Pantheon+SH0ES, resulting in a discrepancy of $2.64\sigma$.

\begin{table}
\centering
\caption{Dilaton Planck and Pantheon+SH0ES Mean Values: A comparison with $\Lambda$CDM.}
\label{tab:dilaton_mean_comparisons}
\begin{tabular}{lccc|ccc}
\toprule 
& \multicolumn{3}{c|}{Planck} & \multicolumn{3}{c}{Pantheon+SH0ES} \\
\cmidrule(lr){2-4} \cmidrule(lr){5-7} & 
$\Lambda$CDM & 
Dilaton & 
Shift &
$\Lambda$CDM & 
Dilaton & 
Shift \\
\midrule
\footnote{The $H_0$ mean values are in ${\rm km s^{-1} Mpc^{-1}}$.}$H_0$ &
$67.27$ & 
$69.57$ & 
$+2.30$ &
$73.37$ & 
$72.70$ & 
$-0.67$ \\

$\Omega_{m0}$ &
$0.317$ & 
$0.325$ & 
$+0.008$ &
$0.335$ & 
$0.342$ & 
$+0.007$ \\

$\Omega_{c0}h^2$ &
$0.1202$ & 
$0.1347$ & 
$+0.0145$ &
$0.1570$ & 
$0.1574$ & 
negligible \\

$\Omega_{b0}h^2$ &
$0.0224$ & 
$0.0224$ & 
negligible &
$0.0231$ & 
$0.0230$ & 
negligible \\
\bottomrule
\end{tabular}
\end{table} 

\begin{figure*}\centering 
\includegraphics[scale=0.55]{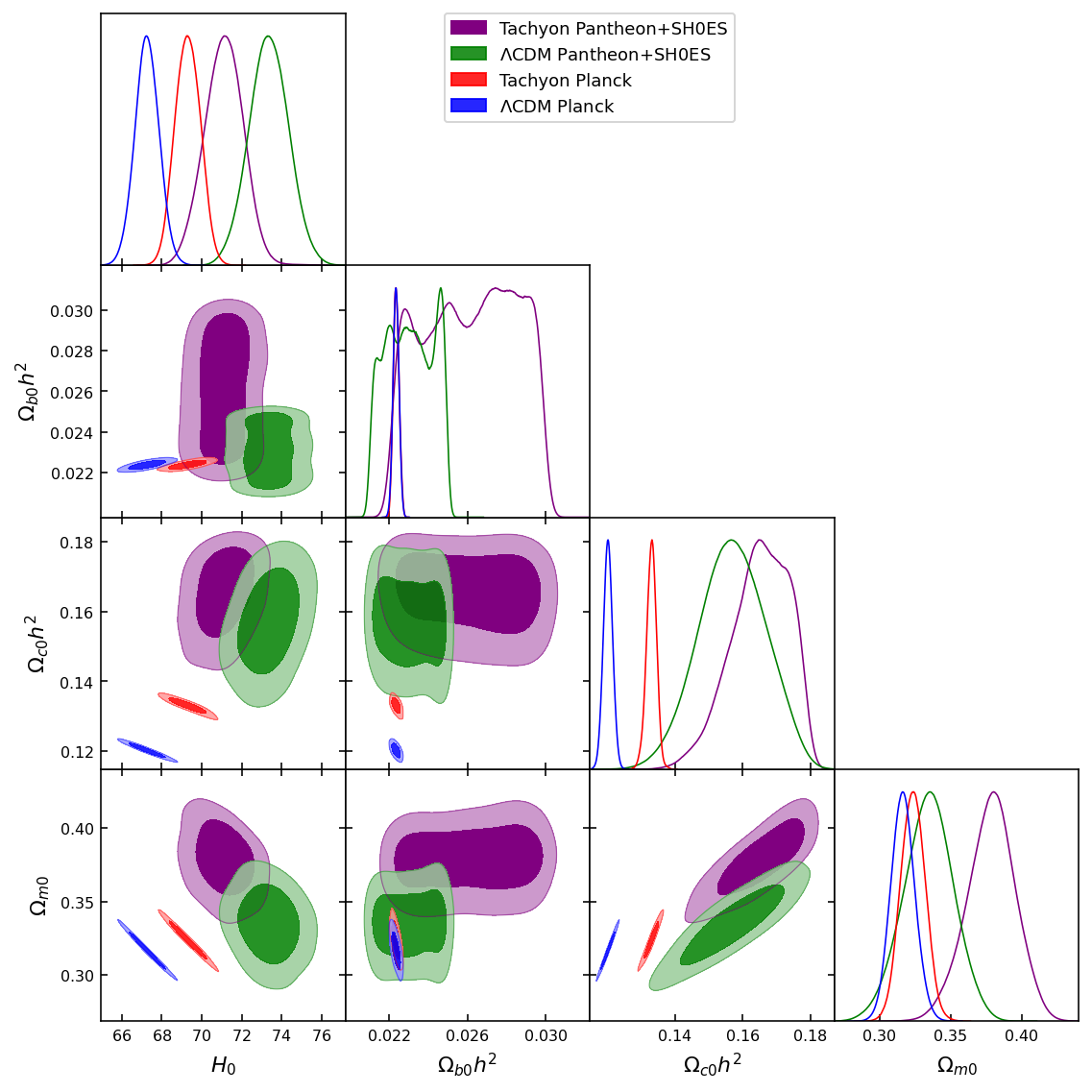}
\caption{\emph{Tachyon}: The contour plots of cosmological parameter estimates with Pantheon+SH0ES and Planck~2018 datasets, for the tachyon and cosmological-constant models. The associated probability density functions are given in the outer panels.}
\label{fig:tachyon-P18-PP}
\end{figure*}

These results imply that the CMB dynamics become modified in the dilaton model, via the sound horizon~\eqref{r_s} (as hinted in~\S\ref{sec:intro}), allowing Planck to accommodate a larger $H_0$ without significantly altering the comoving distance~\eqref{D_M}; consequently, leaving the late-Universe distance ladder constraints (statistically) unchanged. This is a desirable behaviour from a successful Hubble-tension alleviation mechanism: it moves the early-Universe inference toward the late-Universe measurement rather than forcing substantial changes in the supernova-calibrated value. The shift in $\Omega_{c0}h^2$ primarily (and secondarily in $\Omega_{m0}h^2$) gives the compensating cosmological parameter change that permits the increase in the Planck-inferred $H_0$: the best-fit value of $\Omega_{b0}h^2$ (and of model parameters $\lambda \,{\simeq}\, 0.015$ and $A^D_0 \,{\simeq}\, 1.25$) remains largely unchanged. The stability of $\Omega_{b0}h^2$ and the model parameters, between the datasets, therefore implies that the shift in the Planck-preferred $H_0$ in the dilaton model is owing to changes in the dark matter sector rather than modifications of the baryonic physics governing recombination or of the data forcing or favouring a specific k-essence configuration. Thus, in order to explain the Hubble tension between the given probes, the dilaton model will only need to compromise its cold dark matter content (yet preserving its own structure and the baryon density). 

In Fig.~\ref{fig:tachyon-P18-PP} we also show the contour plots of the parameter constraints for the tachyon model and $\Lambda$CDM, using Planck and Pantheon+SH0ES measurements. Similarly, the Planck Hubble-constant posterior exhibits a substantial upward shift. However, unlike in the dilaton scenario, the tachyon also induces a significant shift in all the Pantheon+SH0ES constraints. Again, we summarise the shifts in the best-fit values of the cosmological parameters in the tachyon model relative to $\Lambda$CDM, in Table~\ref{tab:tachyon_mean_comparisons}. The Planck inferences show an upward shift in $H_0$ ($+2.04\, {\rm km s^{-1} Mpc^{-1}}$) with an accompanying up-shift in $\Omega_{c0}h^2$ ($+0.0128$). For the Pantheon+SH0ES values, we get a decrease in $H_0$ ($-2.25\, {\rm km s^{-1} Mpc^{-1}}$) and increases in $\Omega_{c0}h^2$ ($+0.008$) and $\Omega_{b0}h^2$ ($+0.003$). Thus, unlike in the dilaton model where the shift in $H_0$ is achieved through a one-sided convergence (the Planck inference), the shift in the tachyon model is achieved by a two-sided convergence: Planck moving it up and Pantheon+SH0ES simultaneously moving it down. The tachyon therefore does not merely alter the Planck inference, but it also changes the preferred late-time cosmological solution. Moreover, the tachyon model produces a constraint of $H_0 \,{=}\, 69.31 \,{\pm}\, 0.62\,{\rm km\,s^{-1} Mpc^{-1}}$ from Planck and gives $H_0 \,{=}\, 71.12 \,{\pm}\, 0.96\, {\rm km\,s^{-1} Mpc^{-1}}$ from Pantheon+SH0ES, reducing the tension to a discrepancy of only $1.58\sigma$: a near direct agreement (as opposed to $5.31\sigma$ in $\Lambda$CDM). 

\begin{table}
\centering
\caption{Tachyon Planck and Pantheon+SH0ES Mean Values: A comparison with $\Lambda$CDM. Notations are as in Table \ref{tab:dilaton_mean_comparisons}.}
\label{tab:tachyon_mean_comparisons}
\begin{tabular}{lccc|ccc}
\toprule
& \multicolumn{3}{c|}{Planck} 
& \multicolumn{3}{c}{Pantheon+SH0ES} \\

\cmidrule(lr){2-4} \cmidrule(lr){5-7} & 
$\Lambda$CDM & 
Tachyon & 
Shift &
$\Lambda$CDM & 
Tachyon & 
Shift\\
\midrule

$H_0$ &
$67.27$ & 
$69.31$ & 
$+2.04$ &
$73.37$ & 
$71.12$ & 
$-2.25$\\

$\Omega_{m0}$ &
$0.317$ & 
$0.324$ & 
$+0.007$ &
$0.335$ & 
$0.379$ & 
$+0.044$\\

$\Omega_{c0}h^2$ &
$0.1202$ & 
$0.1330$ & 
$+0.0128$ &
$0.157$ & 
$0.165$ & 
$+0.008$\\

$\Omega_{b0}h^2$ &
$0.0224$ & 
$0.0224$ & 
negligible &
$0.0231$ & 
$0.0261$ & 
$+0.003$\\
\bottomrule
\end{tabular}
\end{table}

Quantitatively, the reduction of the Hubble tension between Planck and Pantheon+SH0ES in the dilaton and tachyon models, relative to $\Lambda$CDM, represents reductions of about $50\%$ and $70\%$, respectively. Beyond the shift in $H_0$, the k-essence models also induce characteristic changes in the inferred matter densities. The Planck inference of the cold dark matter density parameter increases from $\Omega_{c0}h^2 \,{=}\, 0.1202 \,{\pm}\, 0.0014$ in $\Lambda$CDM to $\Omega_{c0}h^2 \,{=}\, 0.1347 \,{\pm}\, 0.0015$ and $\Omega_{c0}h^2 \,{=}\, 0.133 \,{\pm}\, 0.0015$ in the dilaton and tachyon models, respectively, while the baryon density remains unchanged. 
In general, although the dilaton model requires an unchanged baryon density and the tachyon model, an enhanced baryon density, relative to $\Lambda$CDM, in order to explain the Pantheon+SH0ES data or late-Universe physics, neither of the k-essence models requires alteration of the baryon density to explain the Planck data or recombination physics. This is important because the baryon density is one of the most robustly measured quantities from the relative heights of the acoustic peaks: both k-essence models preserve this. Overall, the results suggest that k-essence dynamics can naturally, self-mitigate the Hubble tension between Planck and Pantheon+SH0ES. The tachyon model brings the two measurements into near direct agreement (${<}\, 2\sigma$), while the dilaton model reduces the tension to a level (${<}\, 3\sigma$) that is no longer regarded as severe (or falls within the range of statistical error). 

\begin{table*}\centering
\caption{Statistical constraints for $\Lambda$CDM and k-essence models derived from Planck~2018, Pantheon+SH0ES (PP), and different combinations of complementary late-Universe probes with PP. We also give the $H_0$ tensions with Planck.}
\label{tab:combinations}
\begin{tabular}{l|ccccccc|c}
\toprule
\multicolumn{9}{l}{\textbf{Model}\hfill \textbf{Tension with Planck:} } \\
\midrule

Datasets & 
$H_0$ & 
$\Omega_{m0}$ & 
$\Omega_{b0}h^2$ & 
$\Omega_{c0}h^2$ & 
$M_B$ &&& 
$\Delta{H}_0$\\ 

\midrule
\multicolumn{7}{l}{\textbf{$\Lambda$CDM}}&& \\
\hline

Planck & 
$67.27\pm 0.60$ & 
$0.317\pm 0.009$ & 
$0.0224\pm 0.0002$ & 
$0.1202\pm 0.0014$ & 
$-$ & && 
$-$ \\

PP & 
$73.37\pm 0.98$ & 
$0.335\pm 0.018$ & 
$0.0231^{+0.0018}_{-0.0013}$ & 
$0.1570\pm 0.0099$ & 
$-19.25\pm 0.03$ & && 
$5.31\sigma $ \\

PP+CC & 
$70.80\pm 0.76$ & 
$0.278\pm 0.011$ & 
$0.0255\pm 0.0026$ & 
$0.1139\pm 0.0050$ & 
$-19.34\pm 0.02 $& && 
$3.65\sigma$\\

PP+Union3 & 
$73.18\pm 0.95$ & 
$0.342\pm 0.015$ & 
$0.0256\pm 0.0026$ & 
$0.1576\pm 0.0090$ & 
$-19.25\pm 0.03$ & && 
$ 5.26\sigma$ \\

PP+DESY5 & 
$69.60\pm 0.27$ & 
$0.337\pm 0.012$ & 
$0.0255\pm 0.0026$ & 
$0.1379\pm 0.0056$& 
$-19.36\pm 0.01$ & &&
$3.54 \sigma$ \\

PP+DESI & 
$74.52^{+0.51}_{-0.71}$ & 
$0.304^{+0.007}_{-0.008}$ & 
$0.0222^{+0.0004}_{-0.0011}$ &
$0.1469\pm 0.0053$ & 
$-19.223^{+0.016}_{-0.021}$ & &&
$8.47 \sigma$  \\

ALL\footnote{ALL denotes:~PP + CC + Union3 + DESY5 + DESI.} & 
$70.93\pm 0.16$ & 
$0.283\pm 0.005$ & 
$0.0211^{+0.00002}_{-0.00009}$ & 
$0.1213\pm 0.0023$ &
$-19.34\pm 0.01$ & && 
$5.89 \sigma$ \\

\midrule
\multicolumn{5}{l}{\textbf{Dilaton}}&& 
${\lambda}$ &
${A_0^D}$ & \\

\hline

Planck & 
$69.57\pm 0.64$ & 
$0.325\pm 0.009$ & 
$0.0224\pm 0.0002$ & 
$0.1347\pm 0.0015$ & 
$-$ & 
$ 0.015\pm 0.003$ & 
$1.25\pm 0.14$ & 
$ - $ \\

PP & 
$72.70\pm 1.00$ & 
$ 0.342\pm 0.017$ & 
$0.0230\pm 0.0012$ & 
$0.1574\pm 0.0098$ & 
$-19.25\pm 0.03$ & 
$0.015\pm 0.003$ & 
$1.24\pm 0.14$ &
$2.64\sigma $ \\

PP+CC & 
$70.14\pm 0.75$ & 
$0.284\pm 0.012$ & 
$0.0230\pm 0.0012$ & 
$0.1165\pm 0.0045$& 
$-19.34\pm 0.02$ & 
$0.015\pm 0.003$ & 
$1.25\pm 0.14$ & 
$0.58\sigma$\\

PP+Union3 & 
$72.55\pm 0.95$ & 
$0.348\pm 0.015$ & 
$0.0230\pm 0.0011$ & 
$0.1601\pm 0.0089$ & 
$-19.25\pm 0.03$ &
$0.015\pm 0.003$ & 
$1.25\pm 0.14$ & 
$2.60\sigma$\\

PP+DESY5 & 
$68.96^{+0.26}_{-0.29}$ & 
$0.343\pm 0.012$ & 
$0.0254\pm 0.0026$ & 
$0.1377\pm 0.0055$ & 
$-19.36\pm 0.01$ & 
$0.015\pm 0.004$ & 
$1.26\pm 0.15$ & 
$0.88 \sigma$ \\

PP+DESI & 
$72.56^{+0.90}_{-0.65}$ & 
$0.312\pm 0.008$ & 
$0.0281^{+0.0016}_{-0.0008}$ &
$0.1360\pm 0.0052$ & 
$-19.26^{+0.03}_{-0.02}$ & 
$0.015\pm 0.003$ & 
$1.26\pm 0.14$ & 
$ 2.98\sigma$  \\

ALL & 
$69.66\pm 0.16$ & 
$0.318\pm 0.006$ & 
$ 0.0226\pm 0.0007$ & 
$0.1315\pm 0.0026$ &
$-19.36\pm 0.01$ &
$0.011^{+0.004}_{-0.001}$ &
$0.71^{+0.39}_{-0.22}$ &
$0.14\sigma$\\		

\midrule
\multicolumn{5}{l}{\textbf{Tachyon}}&& ${\alpha}$& ${A_0^T}$& \\
\hline

Planck & 
$69.31\pm 0.62$ & 
$0.324\pm 0.009$ & 
$0.0224\pm 0.0002$ & 
$0.1330\pm 0.0015$ &
$-$ &
$0.025\pm 0.003$ &
$1.24\pm 0.43$ & 
$-$ \\

PP & 
$71.12\pm 0.96$ & 
$0.379^{+0.018}_{-0.016}$ & 
$0.0261\pm 0.0023$ & 
$0.165^{+0.011}_{-0.0063}$ & 
$-19.26\pm0.03$ & 
$0.025\pm 0.003$ & 
$1.26\pm 0.43$ &
$1.58\sigma$ \\

PP+CC & 
$68.67\pm 0.76$ & 
$0.308\pm 0.012$ & 
$0.0259^{+0.0024}_{-0.0032}$ & 
$0.1191\pm 0.0048$ & 
$-19.35\pm 0.02$ &
$0.025^{+0.003}_{-0.004}$ &
$1.25\pm 0.43$ &
$ 0.65\sigma$\\

PP+Union3 & 
$71.13^{+0.97}_{-0.88}$ & 
$0.379\pm 0.014$ & 
$0.0262^{+0.0034}_{-0.0016}$ & 
$0.1654^{+0.0092}_{-0.0069}$ & 
$-19.26\pm0.03$ &
$0.025\pm 0.004$ &
$1.24\pm 0.44$ &
$1.63\sigma$ \\

PP+DESY5 & 
$67.92\pm 0.30$ & 
$0.380\pm 0.013$ & 
$0.0259\pm 0.0023$ & 
$0.1493\pm 0.0052$ & 
$-19.36\pm 0.01$ &
$0.025\pm 0.003$ &
$1.24\pm 0.43$ & 
$2.02\sigma$ \\

PP+DESI & 
$71.52\pm 0.90$ & 
$0.323\pm 0.008$ & 
$0.0278^{+0.0013}_{-0.0010}$ &
$0.1373\pm 0.0055$ & 
$-19.26\pm0.03$ &
$0.025\pm 0.003$ &
$1.26\pm 0.43 $ & 
$ 2.02\sigma $\\

ALL & 
$68.85\pm 0.24$ & 
$0.323\pm 0.006$ & 
$0.0246\pm 0.0006$ & 
$0.1283\pm 0.0028$ &
$-19.34\pm 0.01$ &
$0.025\pm 0.003$ &
$1.25\pm 0.43$ & 
$0.69 \sigma$\\
\bottomrule
\end{tabular}
\end{table*}

The different behaviour of the dilaton and tachyon models can be understood in terms of their distinct dynamics (see e.g.~\cite{Duniya:2026aeu}, for the background evolutions). Since the scalar field contributes directly to the Friedmann equation \eqref{Friedmann} through its non-canonical kinetic structure, each k-essence Lagrangian, \eqref{Dilaton} and \eqref{Tachyon}, will produce a different expansion history despite describing the same late-Universe acceleration. Consequently, the CMB distance degeneracy is realised at different combinations of $H_0$, $\Omega_{m0}$ and $\Omega_{c0}h^2$. The dilaton background remains sufficiently close to $\Lambda$CDM expansion history that the late-Universe constraints are only weakly modified, leading mainly to an upward shift of the Planck estimation. On the other hand, the tachyon background departs more noticeably from $\Lambda$CDM, causing both the Planck and late-Universe estimations to move towards a common region of parameter space.



\subsection{Consistency Across Complementary Probes}
\label{subsec:combinations}

As previously stated, the Hubble tension is fundamentally a disagreement between $H_0$ values inferred independently from early- and late-Universe observations. A reduction of the Planck--Pantheon+SH0ES tension alone within a given cosmological model is not a complete resolution of the Hubble tension by the model. Instead, the reduction should persist when tested against complementary probes of the expansion history that employ different observational techniques and are subject to different systematic uncertainties. Thus, we compare Planck with different late-Universe measurement combinations comprising Pantheon+SH0ES (PP), PP+CC, PP+Union3, PP+DESY5, PP+DESI and ALL (the combined compilation, i.e. PP+CC+Union3+DESY5+DESI). 

In Table~\ref{tab:combinations}, we give the full constraints corresponding to the given datasets, for $\Lambda$CDM and the k-essence models. Firstly, we note that across all the different dataset combinations considered, the preferred tachyon parameters remain remarkably stable ($\alpha \,{\simeq}\, 0.025$ and $A_0^T\,{\simeq}\, 1.26$). For the dilaton, except slightly for ALL, the parameters also remain largely unchanged ($\lambda \,{\simeq}\, 0.015$ and $A_0^D\,{\simeq}\, 1.26$). This indicates that the k-essence configuration favoured by the CMB is also compatible with the (BAO-, CC-, and supernova-constrained) late-Universe expansion history. Essentially, the underlying k-essence dynamics are effectively insensitive to the constraints of individual observations. Thus, any changes in the estimations in the k-essence models will arise from the modified expansion history of the k-essence cosmology rather than from an alteration of the observational data or variations in the core k-essence parameters. (See \S\ref{subsec:Planck_PP}, for a dedicated discussion on the Planck and Pantheon+SH0ES results.) 

By combining Pantheon+SH0ES with cosmic chronometer measurements, the resulting PP+CC dataset provides a complementary late-Universe probe by jointly constraining the expansion history through luminosity distances and direct, model-independent determinations of the Hubble parameter $H(z)$. In $\Lambda$CDM, we obtain $H_0 \,{=}\, 70.80\, {\pm}\, 0.76$ ${\rm km\, s^{-1} Mpc^{-1}}$ from the PP+CC joint analysis. This evaluates to a discrepancy of $3.65\sigma$ relative to the Planck predicted value. Moreover, the accompanying changes in the matter and cold dark matter density parameters from the PP+CC joint analysis are decreases, ($\Omega_{m0}$: $0.317\, {\pm}\, 0.009\, {\to}\, 0.278\, {\pm}\, 0.011$) and ($\Omega_{c0}h^2$: $0.1202\, {\pm}\, 0.0014\, {\to}\, 0.1139\, {\pm}\, 0.0050$), respectively, relative to the Planck determinations. The baryon density parameter increases ($\Omega_{b0}h^2$: $0.0224 \,{\pm}\, 0.0002\, {\to}$ $0.0255\, {\pm}\, 0.0026$). Unlike in the Pantheon+SH0ES-only analysis, the inclusion of cosmic chronometer measurements significantly lowers the preferred $H_0$, reducing the Hubble tension but still leaving a significant discrepancy within $\Lambda$CDM. For the dilaton model, the same dataset yields $H_0 \,{=}\, 70.14\, {\pm}$ $0.75$ ${\rm km\, s^{-1} Mpc^{-1}}$, reducing the discrepancy with the corresponding Planck prediction to only $0.58\sigma$. Relative to the Planck inference, the matter density parameter from PP+CC decreases from $\Omega_{m0} \,{=}\, 0.325\, {\pm}\, 0.009$ to $0.284\, {\pm}\, 0.012$, while the cold dark matter density parameter remains close to its Planck inference ($\Omega_{c0}h^2$: $0.1347\, {\pm}\, 0.0015 \,{\to}$ $0.1165\, {\pm}\, 0.0045$). Similarly, the tachyon model gives $H_0 \,{=}\, 68.67\, {\pm}\, 0.76$ ${\rm km\, s^{-1} Mpc^{-1}}$, indicating a direct agreement between  the Pantheon+SH0ES+CC and Planck $H_0$ determinations, at only $0.65\sigma$ difference. The PP+CC joint analysis also gives a modestly reduced matter density parameter ($\Omega_{m0}$: $0.324\, {\pm}\, 0.009\, \,{\to}\, 0.308\, {\pm}\, 0.012$) relative to that of Planck, and a cold dark matter density parameter which remains close to the Planck determination ($\Omega_{c0}h^2$: $0.1330\, {\pm}\, 0.0015 \,{\to}$ $0.1191\, {\pm}\, 0.0048$). 

The results indicate that the excellent $H_0$ agreement with the corresponding Planck prediction, in both k-essence models, arises (as previously stated) largely from the modified expansion history of the k-essence cosmology rather than from any change in the observational data, as the preferred model parameters remain unchanged for Planck and PP+CC; with the changes in the dark matter fraction being the cosmological compromise.  Moreover, the PP+CC joint analysis demonstrates that combining Type~Ia supernova observations with direct measurements of the cosmic expansion rate substantially improves the consistency between early- and late-Universe determinations in the k-essence cosmologies. Although the Hubble tension is reduced but remains significant at $3.65\sigma$ within $\Lambda$CDM, it falls below the $1\sigma$ level in both the dilaton and tachyon models under identical observational constraints. This comparison shows that the level of agreement between Planck and late-Universe probes depends sensitively on the assumed cosmological dynamics, with the k-essence models naturally providing a more consistent description of the combined supernova and cosmic chronometer data without altering the observational measurements themselves.

The PP+Union3 dataset provides a consistency test with two (fairly) independent supernova measurements or luminosity-distance determinations, in the late Universe. The analysis of the PP+Union3 dataset gives $H_0 \,{=}\, 73.18\, {\pm}\, 0.95$ ${\rm km\, s^{-1} Mpc^{-1}}$ in $\Lambda$CDM. This evaluates to a difference of $5.26\sigma$ from the Planck-inferred value, indicating that the addition of Union3 leaves the Hubble tension largely unchanged relative to the PP-only analysis. In comparison to the Planck determination, the PP+Union3 matter density (parameter) shifts upward ($\Omega_{m0}$: $0.317\, {\pm}\, 0.009\, {\to}\, 0.342\, {\pm}\, 0.015$), and the cold dark matter density rises ($\Omega_{c0}h^2$: $0.1202\, {\pm}\, 0.0014\, {\to}$ $0.1576\, {\pm}\, 0.0090$); similarly, for the baryon density ($\Omega_{b0}h^2$: $0.0224\, {\pm}\, 0.0002 \,{\to}\, 0.0256\, {\pm}\, 0.0026$), although with larger uncertainty than the Planck determination. These results indicate that supplementing Pantheon+SH0ES with the Union3 compilation does not significantly modify the late-time expansion rate inferred in $\Lambda$CDM, and therefore does not reduce the discrepancy with the Planck prediction. In the dilaton model, the PP+Union3 joint analysis gives $H_0 \,{=}\, 72.55\, {\pm}\, 0.95$ ${\rm km\, s^{-1} Mpc^{-1}}$, reducing the discrepancy with the corresponding Planck determination to $2.60\sigma$. As in $\Lambda$CDM, the corresponding matter density parameter increases modestly ($\Omega_{m0}$: $0.325\, {\pm}\, 0.009\, {\to}\, 0.348\, {\pm}\, 0.015$); whereas, the cold dark matter density parameter changes only slightly ($\Omega_{c0}h^2$: $0.1347\, {\pm}\, 0.0015 \,{\to}\, 0.1601\, {\pm}\, 0.0089$). We see that, although the Hubble tension is not completely removed, it is reduced by approximately a factor of two compared with the corresponding $\Lambda$CDM tension, under identical observational constraints. 

In the tachyon model, the PP+Union3 dataset leads to a Hubble constant $H_0 \,{=}\, 71.13^{+0.97}_{-0.88}$ ${\rm km\, s^{-1} Mpc^{-1}}$, corresponding to a $1.63\sigma$ discrepancy with its Planck prediction. Similarly, the accompanying changes in the matter density parameters are increases, ($\Omega_{m0}$: $0.324\, {\pm}\, 0.009 \,{\to}$ $0.379\,{\pm}\, 0.014$) and ($\Omega_{c0}h^2$: $0.1330\,{\pm}\, 0.0015\, {\to}\, 0.1654^{+0.0092}_{-0.0069}$). In both k-essence models, the improved agreement with the Planck prediction originates from the modified expansion history of the k-essence cosmology rather than from any alteration of the observational data, with the underlying dark matter content being compromised to compensate for this modification. Overall, the PP+Union3 analysis shows that combining two independent SNIa compilations produces markedly different levels of consistency between the early- and late-Universe determination of the present-day expansion rate depending on the assumed cosmological model. While the inferred Hubble tension remains above $5\sigma$ within $\Lambda$CDM, it is reduced to $2.60\sigma$ in the dilaton model, and to only $1.63\sigma$ in the tachyon model (under the same observational constraints). Similar to the case of PP+CC, k-essence naturally provides a more consistent description of the combined supernova data compared to the cosmological constant. 

Combining Pantheon+SH0ES with DESY5 provides a high-precision late-Universe probe of the expansion history with two complementary supernova datasets. In $\Lambda$CDM, we obtain $H_0 \,{=}\, 69.60 \,{\pm}\, 0.27$ ${\rm km\, s^{-1} Mpc^{-1}}$ from the PP+DESY5 joint dataset. This inference corresponds to a $3.54\sigma$ tension with the Planck predicted value. Both the inferred matter and cold dark matter densities increase, from $\Omega_{m0} \,{=}\, 0.317\, {\pm}\, 0.009$ to $0.337 \,{\pm}\, 0.012$ and from $\Omega_{c0}h^2 \,{=}\, 0.1202 \,{\pm}\, 0.0014$ to $0.1379 \,{\pm}\, 0.0056$, respectively. Similarly, the baryon density shifts upward, from $\Omega_{b0}h^2 \,{=}\, 0.0224 \,{\pm}\, 0.0002$ to $0.0255 \,{\pm}\, 0.0026$. Compared to PP-only analysis, the inclusion of the DESY5 sample lowers the preferred value of the present-day expansion rate, reducing the significant discrepancy with the Planck prediction within $\Lambda$CDM. In the dilaton model, the same dataset leads to $H_0 \,{=}\, 68.96_{-0.29}^{+0.26}$ ${\rm km\, s^{-1} Mpc^{-1}}$, reducing the discrepancy with the corresponding Planck prediction to only $0.88\sigma$, indicating direct agreement between Planck and PP+DESY5. The corresponding matter density increases slightly  ($\Omega_{m0}$: $0.325 \,{\pm}\, 0.009\, {\to}\, 0.343 \,{\pm}\, 0.012$), with that of the cold dark matter density remaining close to its Planck inference ($\Omega_{c0}h^2$: $0.1347\, {\pm}\, 0.0015\, {\to}\, 0.1377 \,{\pm}\, 0.0055$). Similarly, in the tachyon model, the PP+DESY5 dataset produces $H_0 \,{=}\, 67.92 \,{\pm}\, 0.30$ ${\rm km\, s^{-1} Mpc^{-1}}$, which corresponds to a discrepancy of $2.02\sigma$ with the Planck determination, while the corresponding matter density increases ($\Omega_{m0}$: $0.324 \,{\pm}\, 0.009 \,{\to}\, 0.380 \,{\pm}\, 0.013$). Moreover, the cold dark matter density increases more moderately ($\Omega_{c0}h^2$: $0.1330 \,{\pm}\, 0.0015 \,{\to}\, 0.1493 \,{\pm}\, 0.0052$). Although the residual discrepancy is larger than in the dilaton model, it remains significantly smaller than that obtained within $\Lambda$CDM under the same observational constraints. Thus, the PP+DESY5 analysis demonstrates that combining two independent late-time supernova datasets substantially improves the consistency between early- and late-Universe determinations in the k-essence cosmologies, while a significant discrepancy persists within $\Lambda$CDM. Since all three models are fitted to the identical observational data, the differing levels of agreement with the corresponding Planck predictions reflect the distinct cosmological evolution allowed by each model rather than any difference in the underlying measurements.

The combination of Pantheon+SH0ES with DESI provides a stringent test of the Hubble tension by jointly constraining the expansion history through SNIa and BAO. Within $\Lambda$CDM, PP+DESI gives $H_0 \,{=}\, 74.52^{+0.51}_{-0.71}$ ${\rm km\, s^{-1} Mpc^{-1}}$, corresponding to an $8.47\sigma$ tension with the Planck predicted value. This represents the largest discrepancy among all the dataset combinations considered. Also, relative to the Planck prediction, the matter density decreases ($\Omega_{m0}$: $0.317\,{\pm}\,0.009\, {\to}\, 0.304^{+0.007}_{-0.008}$) and the cold dark matter density increases substantially ($\Omega_{c0}h^2$: $0.1202\,{\pm}\, 0.0014\, {\to}\, 0.1469\,{\pm}\, 0.0053$). In contrast, the baryon density remains consistent with the Planck value, $\Omega_{b0}h^2 \,{=}\, 0.0224\,{\pm}\, 0.0002\, {\to}\, 0.0222^{+0.0004}_{-0.0011}$, indicating that the key effect of combining Pantheon+SH0ES with DESI is to favour a significantly higher present-day expansion rate together with an enhanced cold dark matter density. The resulting increase in the tension demonstrates that the inclusion of BAO data alone does not alleviate the Hubble tension in $\Lambda$CDM. For the dilaton model, PP+DESI gives $H_0 \,{=}\, 72.56^{+0.90}_{-0.65}$ ${\rm km\, s^{-1} Mpc^{-1}}$, reducing the discrepancy with the corresponding Planck prediction to $2.98\sigma$. The matter density changes only slightly ($\Omega_{m0}$: $0.325\,{\pm}\,0.009 \,{\to}\, 0.312\,{\pm}\, 0.008$), while the cold dark matter density remains close to its Planck determination ($\Omega_{m0}$: $0.1347\,{\pm}\, 0.0015 \,{\to}\, 0.1360\,{\pm}\, 0.0052$). Thus, the significantly smaller Planck discrepancy compared to $\Lambda$CDM is achieved without requiring substantial changes in the model parameters, but instead follows naturally from the modified cosmological evolution. 

Similarly, the tachyon model gives $H_0 \,{=}\, 71.52\, {\pm}\, 0.90$ ${\rm km\, s^{-1} Mpc^{-1}}$ from the PP+DESI dataset, which corresponds to a discrepancy of $2.02\sigma$ with the Planck prediction. The estimated matter density value remains effectively unchanged ($\Omega_{m0}$:~$0.324\, {\pm}\, 0.009 \,{\to}\, 0.323\, {\pm}\, 0.008$) and the cold dark matter density increases only modestly ($\Omega_{c0}h^2$:~$0.1330\,{\pm}\,0.0015 \,{\to}\, 0.1373\,{\pm}\, 0.0055$). As in the dilaton case, the reduced discrepancy with the Planck prediction therefore reflects the alternative expansion history inherent to the tachyon cosmology rather than any modification of the observational data. Overall, the PP+DESI analysis demonstrates that the interpretation of the same late-time observations depends strongly on the assumed cosmological model. While the addition of DESI substantially amplifies the Hubble tension within $\Lambda$CDM, the corresponding discrepancies remain ${<}\, 3\sigma$ for both k-essence models under identical observational constraints. This comparison shows that the reduced Planck--late-Universe disagreement arises from the different dark sector adjustments in the k-essence models rather than from changes in the data or from fine tuning of model parameters.

\begin{table*}[ht]
\centering
\caption{Shifts $\Delta \equiv (\text{kCDM}) \,{-}\, (\Lambda{\rm CDM})$ in the mean of cosmological parameters in Table~\ref{tab:combinations}. Positive (negative) values indicate that the value in the k-essence model (kCDM) is larger (smaller) than the corresponding $\Lambda$CDM value, for the given dataset.}
\label{tab:combination_shifts}
\begin{tabular}{lccccccc}
\toprule
\multicolumn{8}{l}{\bf Dilaton}\\
\midrule
Parameter \,&\, 
Planck \,&\, 
PP \,&\, 
PP+CC \,&\, 
PP+Union3 \,&\, 
PP+DESY5 \,&\, 
PP+DESI \,&\, 
ALL \\

\midrule

$H_0$ &
$+2.30$ & 
$-0.67$ &
$-0.66$ &
$-0.63$ &
$-0.64$ &
$-1.96$ &
$-1.27$ \\

$\Omega_{m0}$ &
$+0.008$ &
$+0.007$ &
$+0.006$ &
$+0.006$ &
$+0.006$ &
$+0.008$ &
$+0.035$ \\

$\Omega_{c0}h^2$ &
$+0.0145$ &
$+0.0004$ &
$+0.0026$ &
$+0.0025$ &
$-0.0002$ &
$-0.0109$ &
$+0.0102$ \\

$\Omega_{b0}h^2$ &
$0.0000$ &
$-0.0001$ &
$-0.0025$ &
$-0.0026$ &
$-0.0001$ &
$+0.0059$ &
$+0.0015$ \\

\midrule
\multicolumn{8}{l}{\bf Tachyon}\\
\midrule

$H_0$ & 
$+2.04$ &
$-2.25$ &
$-2.13$ & 
$-2.05$ & 
$-1.68$ & 
$-3.00$ & 
$-2.08$ \\

$\Omega_{m0}$ & 
$+0.007$ & 
$+0.044$ & 
$+0.030$ & 
$+0.037$ & 
$+0.043$ & 
$+0.019$ & 
$+0.040$ \\

$\Omega_{c0}h^2$ & 
$+0.0128$ & 
$+0.0080$ & 
$+0.0052$ & 
$+0.0078$ & 
$+0.0114$ & 
$-0.0096$ & 
$+0.0070$ \\

$\Omega_{b0}h^2$ & 
$0.0000$ & 
$+0.0030$ & 
$+0.0004$ & 
$+0.0006$ & 
$+0.0004$ & 
$+0.0056$ & 
$+0.0035$ \\
\bottomrule
\end{tabular}
\end{table*}

The ALL combined dataset provides the most stringent late-Universe test by simultaneously incorporating luminosity-distance measurements, direct expansion-rate observations, updated supernova compilations, and BAO constraints. The ALL joint analysis leads to a present-day expansion rate $H_0 \,{=}\,70.93 \, {\pm}\, 0.16$ ${\rm km\, s^{-1} Mpc^{-1}}$ in $\Lambda$CDM, corresponding to a persistent tension of $5.89\sigma$ with the Planck prediction despite the substantially improved precision. While the corresponding baryon density remains close to the Planck determination, the matter density decreases ($\Omega_{m0}$: $0.317\,{\pm}\, 0.009\, {\to}\, 0.283\,{\pm}\, 0.005$), with that of the cold dark matter remaining largely stable ($\Omega_{c0}h^2$: $0.1202\, {\pm}\, 0.0014\, {\to}\, 0.1213\, {\pm}\, 0.0023$). This indicates that $\Lambda$CDM preserves the quantities that are most tightly constrained by the CMB and primarily alters the inferred present-day expansion rate and matter density in order to fit the combined late-Universe data. The persistence of a statistically significant tension after combining all complementary probes shows that increasing observational precision alone does not reconcile the early- and late-Universe determinations within $\Lambda$CDM. Within the dilaton model, the same combined dataset gives $H_0 \,{=}\, 69.66\, {\pm}\, 0.16$ ${\rm km\, s^{-1} Mpc^{-1}}$, reducing the tension with the corresponding Planck prediction to only $0.14\sigma$. The accompanying change in the matter density remains moderate ($\Omega_{m0}$: $0.325\, {\pm}\, 0.009 \,{\to}\, 0.318\, {\pm}\, 0.006$), with that in the cold dark matter density also remaining small ($\Omega_{c0}h^2$: $0.1347\, {\pm}\, 0.0015 \,{\to}\, 0.1315\, {\pm}\, 0.0026$). At the same time, the preferred dilaton parameters evolve from ($\lambda$, $A_0^D$) = ($0.015\, {\pm}\, 0.003$, $1.25\, {\pm}\, 0.14$) for Planck to ($0.011^{+0.004}_{-0.001}$, $0.71^{+0.39}_{-0.22}$) for the ALL joint dataset, indicating that the data favour a slightly weaker k-essence contribution while retaining the same underlying dynamics. Importantly, the much smaller Planck offset is obtained using identical late-Universe observations as $\Lambda$CDM, indicating that the reduced discrepancy arises from the different cosmological evolution permitted by the dilaton. 

In the tachyon model, $H_0\, {=}\, 68.85\, {\pm}\, 0.24$ ${\rm km\, s^{-1} Mpc^{-1}}$ from the ALL combined dataset analysis, indicating a reduction of the tension with its Planck determination to only $0.69\sigma$ level, while keeping the matter density almost unchanged ($\Omega_{m0}$: $0.324\, {\pm}\, 0.009 \,{\to}\, 0.323\, {\pm}\, 0.006$) and that of cold dark matter only slightly reduced ($\Omega_{c0}h^2$: $0.1330\, {\pm}\, 0.0015 \,{\to}\, 0.1283\, {\pm}\, 0.0028$). Unlike in the dilaton model, the tachyon parameters remain effectively the same between the Planck and the ALL combined analyses ($\alpha \,{\simeq}\, 0.025$, $A_0^T \,{\simeq}\, 1.25$), suggesting that the preferred tachyon dynamics are already strongly constrained by the CMB and remain fully compatible with the late-Universe observations. The substantially smaller Planck offset therefore reflects the different expansion history predicted by the tachyon cosmology under the same observational constraints rather than any modification of the model by the data. The ALL analysis demonstrates that the inferred level of direct agreement between Planck and late-Universe probes depends on the assumed cosmological dynamics. The k-essence models considered consistently infer lower values of $H_0$ relative to their corresponding Planck predictions than in $\Lambda$CDM, with all three models being constrained by the same observations. This demonstrates that the given k-essence cosmologies can substantially alleviate the Hubble tension, by direct test, without altering the observational data or the underlying k-essence itself being deformed by the data, but by providing an alternative description of the expansion history (or dark matter content).


\subsection{Parameter Shifts in Hubble Tension Inference}
\label{subsec:Shifts}

\begin{figure*}\centering 
	\includegraphics[scale=0.5]{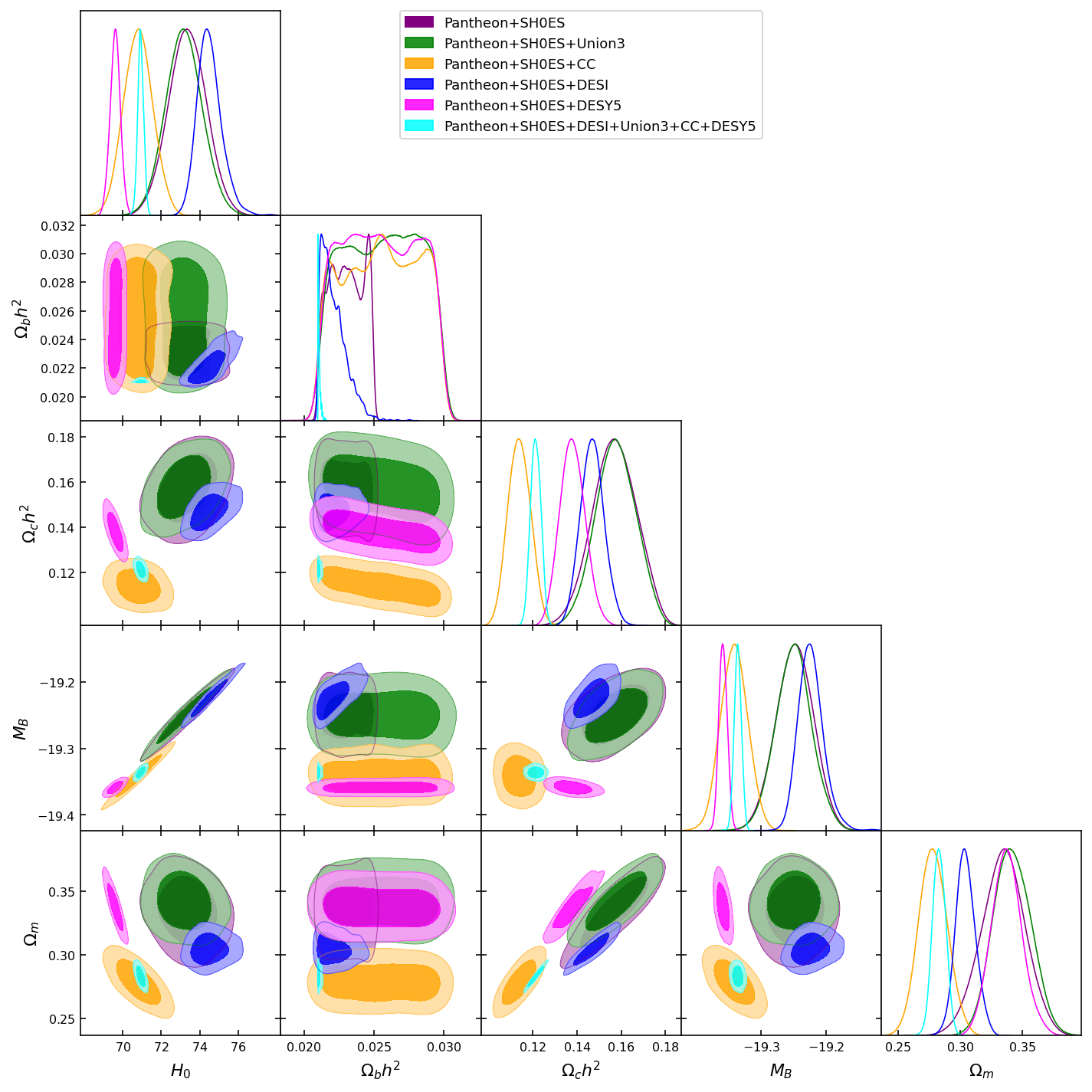}
	\caption{$\Lambda$CDM: Contour plots of cosmological constraints for $\Lambda$CDM derived from Planck~2018, Pantheon+SH0ES, and different combinations of complementary late-Universe measurements with Pantheon+SH0ES.}
	\label{fig:LCDM_combinations}
\end{figure*}

\begin{figure*}\centering 
	\includegraphics[scale=0.5]{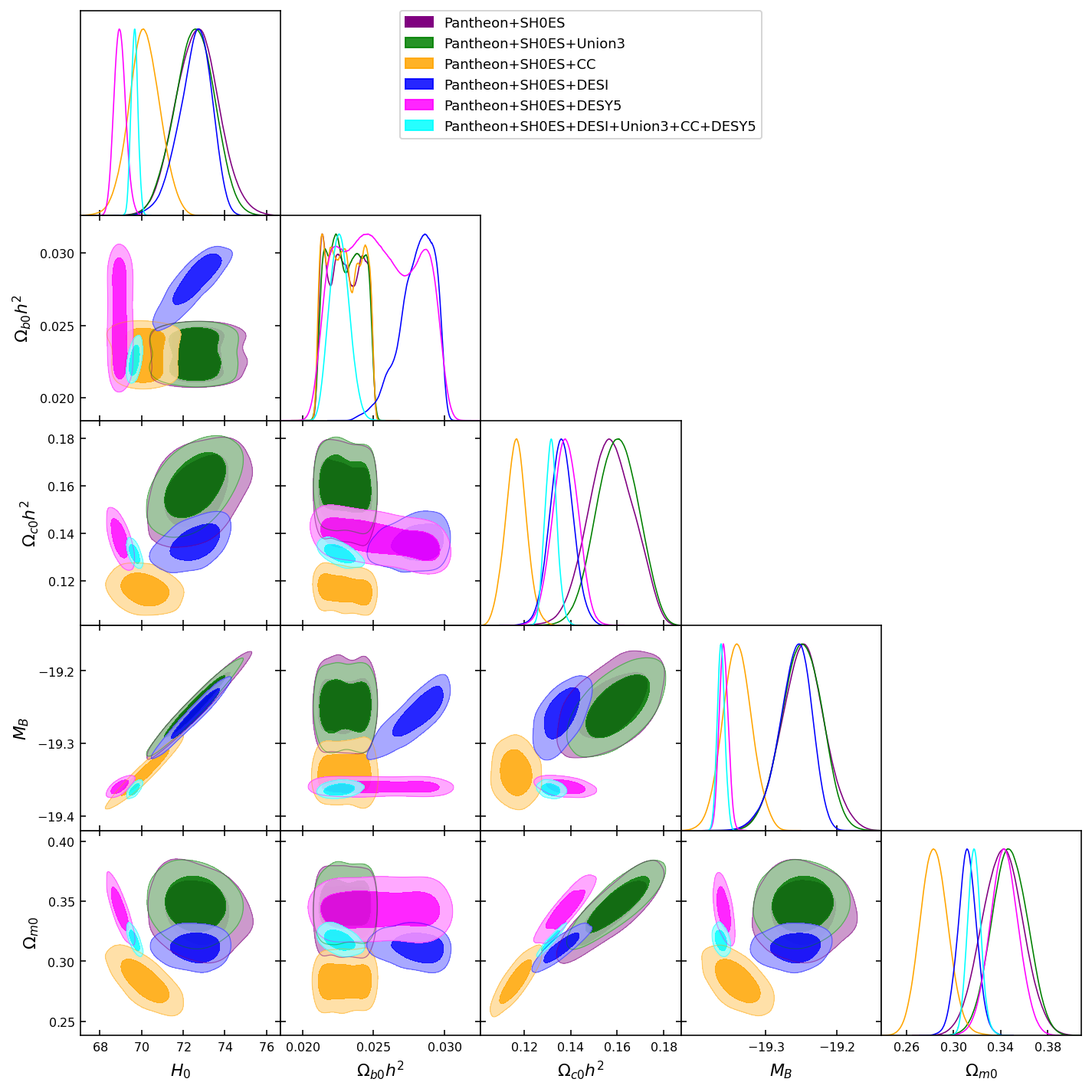}
	\caption{\emph{Dilaton}: Contour plots of cosmological constraints for the dilaton model derived from Planck~2018, Pantheon+SH0ES, and different combinations of complementary late-Universe measurements with Pantheon+SH0ES.}
	\label{fig:dilaton_combinations}
\end{figure*}

\begin{figure*}\centering 
	\includegraphics[scale=0.5]{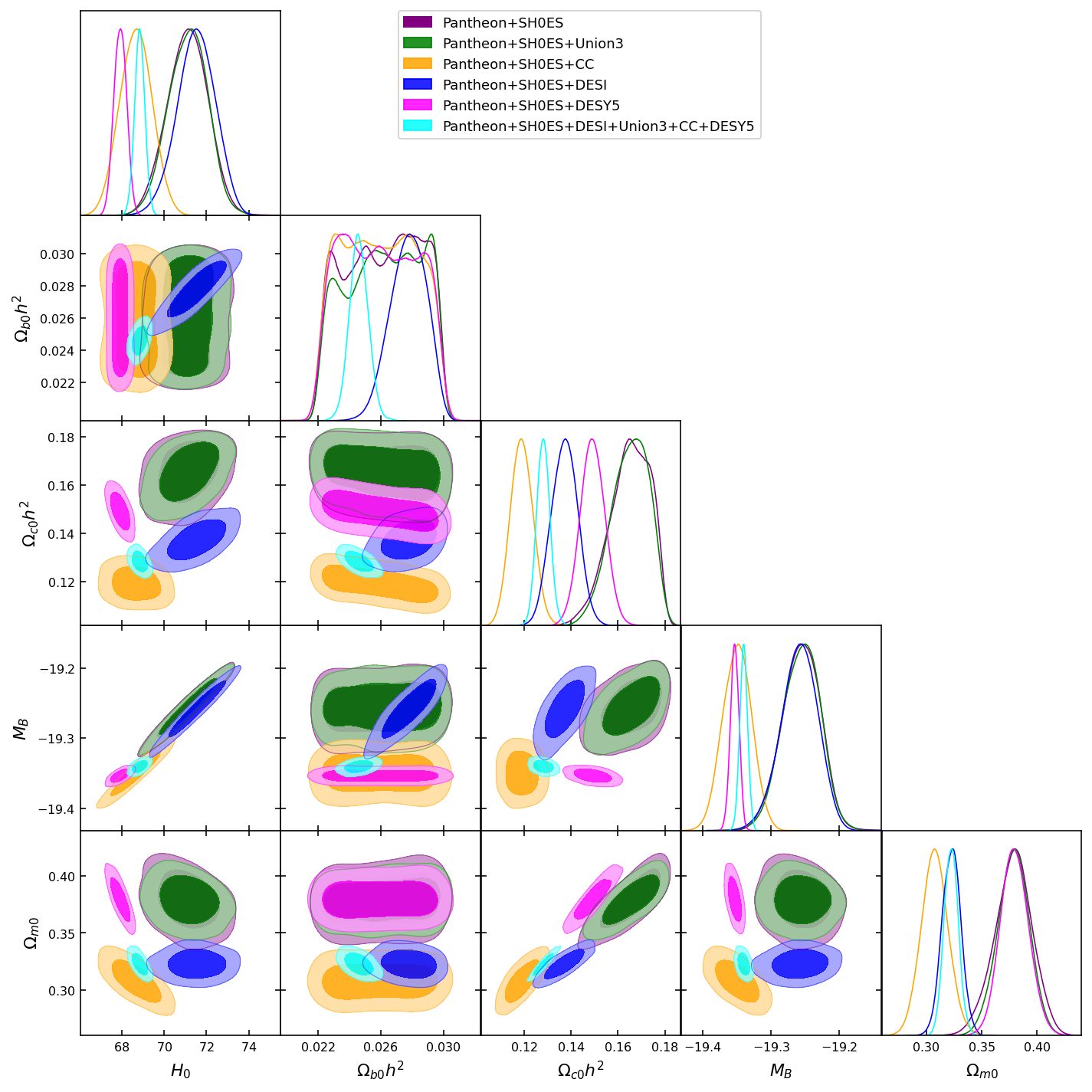}
	\caption{\emph{Tachyon}: Contour plots of cosmological constraints for the tachyon model derived from Planck~2018, Pantheon+SH0ES, and different combinations of complementary late-Universe measurements with Pantheon+SH0ES.}
	\label{fig:tachyon_combinations}
\end{figure*}

We examine the parameter shifts responsible for the Hubble tension alleviation in the k-essence models, relative to $\Lambda$CDM. We give these in Table~\ref{tab:combination_shifts}. (We also give the contours corresponding to Table~\ref{tab:combinations}, in Figs.~\ref{fig:LCDM_combinations} -- \ref{fig:tachyon_combinations}, respectively, for visual intuition.) Table~\ref{tab:combination_shifts} exposes systematic trends that are otherwise obscured in the larger constraint Table~\ref{tab:combinations} and provide quantitative support for interpretation of the distinct behaviours of the dilaton and tachyon models.

In the dilaton model, the dilaton field induces only small changes in the inferred late-Universe cosmological parameters. For the combinations PP+CC, PP+Union3 and PP+DESY5, the Hubble constant decreases by only $\Delta H_0 \,{\simeq}\, {-}(0.63\text{--}0.66)$ ${\rm km\,s^{-1} Mpc^{-1}}$, with the matter density shifting by $\Delta\Omega_{m0} \,{\simeq}\, {+}0.006$. Similarly, the shifts in the cold dark matter density remain within $\left|\Delta(\Omega_{c0}h^2)\right| \,{\lesssim}\, 3\times 10^{-3}$, while the baryon density changes are (statistically) negligible. This stability is notable and implies that, for the given complementary datasets, the dilaton field does not modify the preferred late-Universe Hubble expansion. Instead, it acts to preserve the same expansion history inferred from the late-Universe observations while allowing only modest adjustments in the matter density parameters. The principal exception is the PP+DESI (consequently, ALL) combination, where the preferred value of $H_0$ decreases more substantially ($\Delta H_0 \,{=}\, {-}1.96$ ${\rm km\,s^{-1} Mpc^{-1}}$), accompanied by reductions in both $\Omega_{c0}h^2$ and increases in $\Omega_{b0}h^2$, suggesting that DESI measurements impose stronger constraints on the late-Universe parameter degeneracies. Nevertheless, even in this case the shifts remain considerably smaller than those for the tachyon model. The analysis for ALL exhibits a larger increase in the matter density ($\Delta\Omega_{m0} \,{=}\, {+}0.035$) together with an increase in $\Omega_{c0}h^2$, reflecting the increased constraining power obtained when all complementary probes are analysed simultaneously. Importantly, these shifts remain moderate while the tension between Planck and the late-Universe probes is reduced to only $0.14\sigma$, indicating that the dilaton achieves near-complete consistency without requiring large distortions of the late-Universe parameter space. 

The tachyon model exhibits a qualitatively different behaviour. Across all late-Universe complementary dataset combinations, the shifts are substantially larger and highly coherent. The inferred Hubble constant decreases by approximately $\Delta H_0\, {=}\, {-}(1.7\text{--}3.0)\, {\rm km\,s^{-1} Mpc^{-1}}$, with the largest shift occurring for PP+DESI ($\Delta H_0 \,{=}\, {-}3.0$ $ {\rm km\,s^{-1} Mpc^{-1}}$). Simultaneously, the matter density increases significantly, $\Delta\Omega_{m0} \,{=}\, 0.019\text{--}0.043$, representing increases that are typically five to seven times larger than those found for the dilaton model. The cold dark matter density also increases consistently for all dataset combinations except PP+DESI, where both k-essence models instead favour a lower $\Omega_{c0}h^2$. In contrast to the dilaton model, the tachyon scenario consistently predicts positive shifts in the baryon density, although these remain comparatively small. The coherence of these parameter shifts suggests that the tachyon field does not merely modify the $\Lambda$CDM expansion but instead favours a different late-Universe cosmological parameter space. The simultaneous decrease in $H_0$ and increase in $\Omega_{m0}$ across independent late-Universe datasets indicates that the model systematically rebalances the expansion history and matter content required to fit the observations. The particularly large shifts obtained for PP+DESI further imply that BAO measurements are especially sensitive to the modified expansion dynamics induced by the tachyon field. Additionally, given that the model parameters are not dataset- or observation-dependent, the resulting one-sided convergence in the dilaton case and two-sided convergence in the tachyon case therefore arise naturally from the distinct background evolution of the given k-essence models rather than from differences in the observational datasets themselves.

As already noted in the introduction, and throughout the paper, the present analysis primarily focused on the Hubble tension between Planck and combinations of late-Universe complementary observations, hence discussions and interpretations were restricted appropriately. However, while the present analysis did not include additional early-Universe probes, e.g.~WMAP \cite{WMAP:2012nax}, ACT \cite{ACT:2020gnv} or SPT \cite{SPT-3G:2021eoc}, Planck remains the most precise early-Universe measurement currently available and serves as the gold-standard early-Universe probe in cosmology. Moreover, the Hubble tension is traditionally between Planck and SH0ES (or local $H_0$) measurements. (An extensive study of k-essence using additional early-Universe datasets is deferred to future work.) We demonstrated here that across every of the late-Universe dataset considered (PP, PP+CC, PP+Union3, PP+DESY5, PP+DESI, and ALL), both k-essence models produced a smaller Planck--late-Universe discrepancy than $\Lambda$CDM; that trend was consistent across the independent observational measurements, including supernovae, cosmic chronometers, and BAO. For the most comprehensive dataset, ALL, the Hubble tension drastically reduced from $5.89\sigma$ in $\Lambda$CDM to $0.14\sigma$ for dilaton and $0.69\sigma$ for tachyon; thereby resolving the Planck--late-Universe Hubble tension in the given cosmologies. Moreover, the k-essence parameters were generally stable across datasets, suggesting that the models are not relying on finely tuned parameter shifts to achieve this behaviour. Our results have therefore established a substantial and robust direct alleviation of the Planck-Pantheon+SH0ES Hubble tension (representative of the original Planck--SH0ES tension).


\section{Conclusion}\label{sec:conclusion}	
We presented a detailed analysis of the Hubble tension, in a homogeneous and isotropic background universe consisting photons, baryons, cold dark matter, and k-essence dark energy. We first reevaluate the standard Hubble  tension between the Planck and the SH0ES-calibrated Pantheon+ (Pantheon+SH0ES) datasets, in $\Lambda$CDM and two physically motivated k-essence cosmologies: the dilaton and tachyon models. We further performed successive evaluations of Planck and late-Universe constraints on the Hubble constant, for the same cosmological models, using identical combinations of complementary late-Universe measurements including Union3, DESY5, DESI, and CC. We note that while the present analysis did not include additional early-Universe probes, Planck remains the most precise CMB measurement currently available and serves as the gold-standard early-Universe cosmological probe.

By contrasting the Planck prediction with the corresponding Pantheon+SH0ES determination, we demonstrated that k-essence alleviates the Planck--Pantheon+SH0ES Hubble tension (representative of the original Planck-SH0ES tension), reducing it from $5.31\sigma$ in $\Lambda$CDM to $2.64\sigma$ in the dilaton model and to only $1.58\sigma$ in the tachyon model, with the dark matter density being the cosmological parameter altered by the k-essence dynamics to accommodate the modified expansion history. This reduction in inferred tension between Planck and Pantheon+SH0ES corresponds to reductions of about $50\%$ and $70\%$ in the dilaton and tachyon models, respectively, relative to $\Lambda$CDM. This indicates that k-essence holds the potential to naturally alleviate the tension without any external mediation.

Moreover, we also demonstrated that both k-essence models drastically reduce the Hubble tension between Planck and every late-Universe dataset combination considered (Pantheon+SH0ES with each of Union3, DESY5, DESI, and CC; and the composite of all) to a level well below that of $\Lambda$CDM. The analyses demonstrated that the reduction of the Planck--late-Universe Hubble tension is robust against the inclusion of independent cosmological probes spanning different redshift ranges and observational techniques. Both k-essence models consistently reduced the tension for every dataset combination considered, irrespective of whether the late-Universe constraints are provided by supernovae alone or in combination with cosmic chronometers and BAO measurements, with the combined (late-Universe) joint analysis giving residual offsets of only $0.14\sigma$ and $0.69\sigma$ for the dilaton and tachyon models, respectively, as opposed to $\Lambda$CDM which gave inconsistent tension inferences, yielding a $5.89\sigma$ tension for the joint late-Universe dataset. This established a resolution of the Hubble tension between Planck and late-Universe probes within the k-essence cosmologies. However, to resolve the Hubble tension between the early Universe and the late Universe in general, other early Universe measurements in addition to Planck will need to be considered.

Despite the common outcome, both k-essence models achieved the Planck--late-Universe tension resolution through distinct dynamical or cosmological-parameter responses. The dilaton model preserved the late-Universe cosmological parameter space with only modest adjustments, whereas the tachyon model systematically shifts the inferred early-Universe expansion to higher $H_0$, and the preferred late-Universe toward lower $H_0$ and adjusted matter density. The persistence of these behaviours across complementary probes strengthens the interpretation that the inferred cosmological-parameter shifts are stable, intrinsic consequences of the underlying k-essence dynamics rather than artefacts of a particular observational probe: the model parameters are not dataset-dependent. Within the scope of the datasets analysed, these results therefore provide strong evidence that both k-essence models offer a robust and self-consistent mechanism for alleviating or resolving the general early--late-Universe Hubble tension. 

Furthermore, our results provide evidence that the severity of the Hubble tension is fundamentally model dependent. Although future observations will determine whether these kinds of k-essence dynamics describe cosmic acceleration, our results demonstrate that the apparent Hubble tension is not an unavoidable feature of late-Universe cosmology but depends decisively on the assumed description of dark energy. Additionally, our analysis indicate both k-essence models shift the Planck $H_0$ prediction while (largely) preserving the baryon density. Thus, indicating that the shift in the Planck-predicted $H_0$ is not owing to modifications of the baryonic physics governing recombination. This is important because the baryon density is one of the best measured quantities from the relative heights of the acoustic peaks. 

Moreover, the distinct behaviours of the dilaton and tachyon models provide evidence for two different pathways towards reducing the Hubble tension: the dilaton achieves improved consistency through relatively small modifications of the early-Universe solution or cosmological parameter space, whereas the tachyon does so through a more substantial reconfiguration of both the early- and late-Universe parameter spaces.

\begin{acknowledgments}
We thank Bikarsh Dinda for useful discussion at the initial stage of this work. IO received funding from the German Academic Exchange Service (DAAD) through the In-Country/In-Region Scholarship Program. 
\end{acknowledgments}



\bibliography{Hubble_tension_in_kessence-prd}

\end{document}